\documentclass[11pt]{article}
\pdfoutput=1
\usepackage{jcappub,natbib}
\def\la{\mathrel{\raise.3ex\hbox{$<$\kern-.75em\lower1ex\hbox{$\sim$}}}}

\title{A flashing beacon in axion inflation: recurring bursts of gravitational waves in the strong backreaction regime
} 

\author[a]{Juan Garc\'\i a-Bellido,}
\author[b]{Alexandros Papageorgiou,}
\author[c,d]{Marco Peloso,}
\author[e]{and Lorenzo Sorbo}

\affiliation[a]{Instituto de F\'\i sica Te\'orica UAM/CSIC, Universidad Aut\'onoma de Madrid, Cantoblanco, 28049 Madrid, Spain}
\affiliation[b]{Particle Theory  and Cosmology Group, Center for Theoretical Physics of the Universe, 
Institute for Basic Science (IBS), 34126 Daejeon, Korea}
\affiliation[c]{Dipartimento di Fisica e Astronomia ``Galileo Galilei'', Universit\`a di Padova, 35131 Padova, Italy}
\affiliation[d]{INFN, Sezione di Padova, 35131 Padova, Italy}
\affiliation[e]{Amherst Center for Fundamental Interactions, Department of Physics,
University of Massachusetts, Amherst, MA 01003, U.S.A.}

\abstract{The coupling between a pseudo-scalar inflaton and a gauge field leads to an amount of additional density perturbations and gravitational waves (GWs) that is strongly sensitive to the inflaton speed. This naturally results in enhanced  GWs at (relatively) small scales that exited the horizon well after the CMB ones, and that can be probed by a variety of GW observatories (from pulsar timing arrays, to astrometry, to space-borne and ground-based interferometers). This production occurs in a regime in which the  gauge field significantly backreacts on the inflaton motion. Contrary to earlier assumptions, it has been recently shown that this regime is characterized by an oscillatory behavior of the inflaton speed, with a period of~${\rm O } \left( 5 \right)$ e-folds. Bursts of GWs are produced at the maxima of the speed, imprinting nearly periodic bumps in the frequency-dependent spectrum of GWs produced during inflation. This can potentially generate correlated peaks appearing in the same or in different GWs experiments.}

\begin{document}

\begin{flushright}   CTPU-PTC-23-07  \end{flushright}

\maketitle
\flushbottom

\section{Introduction}
\label{sec:intro} 

One of the standard predictions of inflation is the generation of a slightly red, unpolarized, almost exactly Gaussian component of the stochastic gravitational wave background (SGWB). The amplitude of this component can span tens of orders of magnitude, depending on the energy scale at which inflation took place. As a consequence, the observation of the signature that this background leaves on the B-modes of the Cosmic Microwave Background (CMB) might well be behind the corner. Or, on the other hand, the amplitude of this background might be so small that it will never be observed. Moreover, the red tilt of its spectrum plus the current limit at CMB scales \cite{BICEP:2021xfz}, implies that this SGWB is below the reach of  current (LIGO, Virgo, KAGRA) or upcoming (LISA, Einstein Telescope, Cosmic Explorer) laser interferometers.

The above predictions assume the zero mode of the inflaton(s) to be the only relevant degree(s) of freedom during inflation, so that the only gravitational waves (GWs) generated during inflation are those related to the amplification of the vacuum modes of the metric due to the inflationary expansion.  If, on the other hand, the inflationary Universe contains additional fields with sufficiently large finite-wavelength excitations, then these excitations provide an additional source of GWs, which might be blue and/or parity-odd and/or non-Gaussian, and whose amplitude will generally be not solely determined by the inflationary energy scale.

A simple and well motivated model where a spectrum of gravitational waves with all these properties is generated is characterized by an axion-like inflaton $\phi$ in interaction with a $U(1)$ gauge field through the $\phi F_{\mu\nu}\tilde{F}^{\mu\nu}$ coupling. In this case the rolling inflaton sources the modes of the gauge field of a given helicity, which are amplified proportionally to the quantity $\xi\propto\dot\phi/H$, where $\phi$ is the inflaton field and $H$ is the inflationary Hubble parameter. In the simplest case of constant or adiabatically evolving $\xi$, the amount of gauge field is exponentially sensitive to this parameter, so that a large amplification of gauge modes is easily achieved. Those fields, in their turn, source an additional population of scalar and tensor modes \cite{Barnaby:2010vf,Sorbo:2011rz}. 

Given that the quantity $\xi$ generally increases during inflation, this effect is typically more relevant at the later stages of inflation, when smaller scales than the CMB ones leave the horizon -- that is, this sourced component is generally expected to have a {\em blue spectrum}. This can naturally lead to a situation where, while no effects of these sourced signals are observed at CMB scale, a stochastic background of gravitational waves might be observable by pulsar timing arrays, Gaia astrometry or laser interferometers \cite{Cook:2011hg,Barnaby:2011qe,Domcke:2016bkh,Moore:2017ity,Garcia-Bellido:2021zgu}, while density perturbations might be large enough as to seed a sizeable amount of primordial black holes~\cite{Linde:2012bt,Bugaev:2013fya,Garcia-Bellido:2016dkw,Garcia-Bellido:2017aan,Garcia-Bellido:2017fdg,Ozsoy:2023ryl}.

The fact that the energy in the amplified gauge modes is {\em exponentially} large in $\xi$ implies that during the later stages of inflation, when observable tensors can be produced, these vectors strongly backreact on the inflaton evolution. Such a regime of strong backreaction has first been studied analytically in~\cite{Anber:2009ua}, where it was found that the transfer of energy to gauge fields slows down the rolling of the inflaton, leading to longer inflation and to a plateau in the energy in gauge modes as a function of the number of e-folds.  In~\cite{Barnaby:2011qe}, this was then shown to imply that also the spectrum of produced tensors would plateau at interferometer frequencies.

The analytical study in~\cite{Anber:2009ua} follows the conventional assumption made on models of warm inflation \cite{Berera:1995ie}, where the dissipation of energy from the inflaton, rather that the expansion of the universe, is the main source of friction for the inflaton motion. As always done for this class of models, it was assumed that the inflaton velocity was evolving slowly enough that at any given time $t$ during inflation the spectrum of photons depended only on $\xi(t)$. In other words, it was assumed that the photon spectrum did not retain any memory of the previous evolution of the system. Later on, refs.~\cite{Cheng:2015oqa,Notari:2016npn,Sobol:2019xls,DallAgata:2019yrr,Domcke:2020zez,Gorbar:2021rlt,Durrer:2023rhc} studied the coupled inflaton/gauge field system {\em numerically}, solving the equation of the gauge field mode by mode, effectively accounting for memory effects in the gauge field spectrum. All those works found that, once these effects are accounted for, in the strong backreaction regime $\xi$ performs large oscillations with an ${\cal O}(1)$ fractional amplitude around the value originally derived in~\cite{Anber:2009ua}. While these numerical simulations assumed a homogeneous inflaton, the work~\cite{Caravano:2022epk} solved the full system of inflaton and gauge field excitations on a lattice, also obtaining an oscillatory evolution in excellent agreement with~\cite{Cheng:2015oqa,Notari:2016npn,Sobol:2019xls,DallAgata:2019yrr,Domcke:2020zez,Gorbar:2021rlt,Durrer:2023rhc}. 

More recently, Ref.~\cite{Peloso:2022ovc} obtained the same behavior analytically, in the regime in which the amplitude of these oscillations can be treated as a perturbatively small quantity. The period of those oscillations was found in~\cite{Peloso:2022ovc} to be of approximately $5$ e-folds, which is consistent with the numerical results. The analytical study explicitly shows~\cite{Peloso:2022ovc} that the cause of the instability is a delay between the moment that a given mode is amplified and the moment in which it is relevant for backreaction, as originally argued in~\cite{Domcke:2020zez}, thus causing the memory effect on $\xi \left( t \right)$ that we have mentioned.  

In the present work we study the shape of the spectra of tensors produced by the gauge field modes once those oscillations are accounted for. To obtain those spectra we solve numerically the coupled  system describing the axion/gauge field dynamics.  Similarly to~\cite{Cheng:2015oqa,Notari:2016npn,Sobol:2019xls,DallAgata:2019yrr,Domcke:2020zez,Gorbar:2021rlt,Durrer:2023rhc}, our only approximation is that we neglect the spatial fluctuations in the inflaton field. An argument in favor of the validity of this approximation is the fact that the numerical results of~\cite{Caravano:2022epk}, which were obtained without assuming spatial homogeneity of the inflaton, are in qualitative agreement with those of~\cite{Cheng:2015oqa,Notari:2016npn,Sobol:2019xls,DallAgata:2019yrr,Domcke:2020zez,Gorbar:2021rlt,Durrer:2023rhc} and with those we obtain in this paper.  Our numerical scheme uses an improved treatment of the ultraviolet gauge field modes, so that we are able to carry the numerical integration for the full duration of inflation, and produce results in qualitative agreement with those of the earlier literature, where the comparison can be made.

Moving to the main focus of our paper, the generation of gravitational waves, it is of course safe to neglect their backreaction during inflation. Therefore, we compute their spectra applying the Green function method to the photon source that we have evaluated numerically. It is worth noting that, while the general expression of the graviton Green function depends on the momentum and background expansion history, the large-scale expression of the Green function depends only on the expansion history, but not on the momentum.

The formulae presented in our paper can be applied to any inflationary potential, and the precise quantitative results strongly depend on the potential. However, our main general result is that the tensor power spectrum can show oscillations with an amplitude of several orders of magnitude and period of a couple of decades in frequency. While those oscillations occur around an average spectrum approximately given by the one obtained using the analytical formulae of~\cite{Anber:2009ua,Barnaby:2011qe}, their amplitude and period are large enough to appear as observable features within the bandwidth of various GWs observatories, from pulsar timing arrays, to astrometry, to interferometers. Remarkably, also the difference between the amplitude of left- and right-handed gravitational waves turns out to be frequency dependent, with a ratio of amplitudes ranging from $O(1)$ to five orders of magnitude.

Rather than performing a systematic study of models of inflation, we  discuss just a single example of an inflationary potential that allows to satisfy constraints from CMB and from LIGO-Virgo-KAGRA, while leading to a spectrum of gravitational waves that might be observable by several other forthcoming and future experiments. This should be seen as an existence proof, as other potentials and values of the inflaton-gauge field coupling would lead to significantly different phenomenology.

Our work is organized as follows. In Section~\ref{sec:equations} we present the equations governing our system, that we compare to the case of a constant $\xi$ in Appendix~\ref{app:constxi}. Section~\ref{sec:numerics} contains an overall description of the numerical techniques used to solve the equations presented in the previous section. The details of the numerical implementation are discussed in Appendices~\ref{app:codevar}, \ref{app:numerics1} and \ref{app:numerics2}. We present the results of our numerical study in Section~\ref{sec:results} and Appendix~\ref{app:parity}, and we summarize and conclude in Section~\ref{sec:conclusions}.

\section{An axionic inflaton interacting with a $U(1)$ field, and generation of tensor modes} 
\label{sec:equations}

In this section we present the equations controlling the evolution of the system. In Subsection \ref{subsec:eq-phi-A} we present the equations for the axion-inflaton and gauge fields. In Subsection \ref{subsec:eq-GW} we then provide the expressions controlling the production of the tensor modes from gauge fields. Contrary to most of the existing literature, that provides these expressions in terms of specific solutions for the gauge modes (most often, with constant parameter $\xi$), we present these relations in terms of arbitrary gauge field mode functions, in which we later insert the numerical solutions obtained from the system of Subsection \ref{subsec:eq-phi-A}.  In Subsection \ref{subsec:Green} we then formally solve the equation controlling the GW production, paying particular attention to the Green function employed in the solution. Within each subsection the reader is referred to specific equations in Appendix \ref{app:codevar}, where the main equations obtained in that subsection are rewritten in term of the variables used in our numerical integrations. Moreover, to make contact with the literature, in Appendix  \ref{app:constxi} we evaluate some of these results for constant $\xi$ (defined in eq.~(\ref{def-xi})).

\subsection{Equations for the inflaton and the gauge fields}
\label{subsec:eq-phi-A}

The model is defined by the action 
\begin{equation}
S = \int d^4 x \, \sqrt{-g} \left[ \frac{M_p^2}{2} \, R - \frac{1}{2} \left( \partial \phi \right)^2 - V \left( \phi \right) - \frac{1}{4} F^2 - \frac{\phi}{4f} \, {\tilde F} \, F \right] \;, 
\end{equation} 
where $F_{\mu \nu} \equiv \partial_\mu A_\nu - \partial_\nu A_\mu$ is the field strength of a U(1) gauge field, and ${\tilde F}^{\mu \nu} \equiv \frac{\eta^{\mu \nu \alpha \beta} F_{\alpha \beta}}{2\,\sqrt{-g}}$ its dual, with $\eta^{\mu \nu \alpha \beta}$ totally antisymmetric and $\eta^{0123} = 1$. The FLRW line element is given by $ds^2 = - d t^2 + a^2 \left( t \right) d \vec{x}^2 = a^2 \left( \tau \right) \left[ - d \tau^2 + d \vec{x}^2 \right]$.  

For a homogeneous inflaton we are free to choose the gauge $A_0 = \vec{\nabla} \cdot \vec{A} = 0$, in which one obtains the equations \cite{Barnaby:2011vw} 
\begin{eqnarray}
&& \vec{A} - \nabla^2 \vec{A} - \frac{\phi'}{f} \, \vec{\nabla} \times \vec{A} = 0  \;, \nonumber\\ 
&& \phi'' + 2 {\cal H} \phi' + a^2 \, \frac{d V}{d \phi} = \frac{a^2}{f} \, \vec{E} \cdot \vec{B} \;, \nonumber\\ 
&& {\cal H}^2 = \frac{1}{3 M_p^2} \left[ \frac{1}{2} \phi^{'2} +  a^2 \, V + \frac{a^2}{2} \left( \vec{E}^2 + \vec{B}^2 \right) \right] \;, 
\label{eom-phi-A}
\end{eqnarray}
where ${\cal H} \equiv \frac{a'}{a}$, while prime denotes derivative with respect to the conformal time $\tau$. Although the gauge field is not necessarily identified with the Standard Model photon, for convenience we have used electromagnetic notation, with~\footnote{To simplify the notation, we omit the symbol hat indicating an operator in presence of the vector sign.}
\begin{equation}
\vec{E} = - \frac{1}{a^2} \, \vec{A}' \;\;,\;\; \vec{B} = \frac{1}{a^2} \, \vec{\nabla} \times \vec{A} \;. 
\end{equation} 

We decompose the gauge field as
\begin{equation}
{\hat A}_i ( \tau ,\, \vec{x} ) = \int \frac{d^3 k}{\left( 2 \pi \right)^{3/2}} \, {\rm e}^{ i\vec{k} \cdot \vec{x}} \, \hat{A}_i ( \tau ,\, \vec{k} ) = \sum_{\lambda = \pm} \int \frac{d^3 k}{\left( 2 \pi \right)^{3/2}} \left[ \epsilon_i^{(\lambda)} ( \vec{k} ) A_\lambda ( \tau ,\, k ) {\hat a}_\lambda ( \vec{k} ) \, {\rm e}^{ i\vec{k} \cdot \vec{x}} + {\rm h.c.} \right] \;, 
\label{A-deco}
\end{equation} 
where $\left[ \hat{a}_\lambda ( \vec{k} ) ,\,  {\hat a}_\sigma^\dagger \left( \vec{p} \right) \right] = \delta_{\lambda \sigma} \, \delta^{(3)} ( \vec{k} - \vec{p} )$, while the properties of the polarizations operators are given in ref.~\cite{Barnaby:2011vw}, with $\lambda = +1$ (respectively, $\lambda = -1$) corresponding to the left-handed (respectively, right-handed) circular polarization. The gauge field mode functions satisfy 
\begin{equation}
A_\pm'' + \left( k^2 \mp k \, \frac{\phi'}{f} \right) A_\pm = 0 \;. 
\label{eq-Ak}
\end{equation} 
This expression often appears in the literature as 
\begin{equation}
A_\pm'' + \left( k^2 \mp 2 \xi a H k \right) A_\pm = 0 
\;,\hspace{1cm} \xi \equiv \frac{\dot{\phi}}{2 f H} \;,  
\label{def-xi}
\end{equation} 
where dot denotes derivative with respect to the physical time $t$. For a slow-roll inflationary solution, the quantity $\xi$ is constant to first order in slow roll. The gauge field amplification is in this case exponentially sensitive to $\xi$, and this combination is the most immediate quantity that allows to asses the amount of production (as a rough estimate, visible signatures at CMB scales require $\xi \simeq 2.5$, while signatures at smaller scales typically occur for $\xi \simeq 5$, see ref.~\cite{Peloso:2016gqs}). In this work we explore the production in the case of varying $\xi$, as it occurs in the regime of significant backreaction of the gauge fields on the background inflaton evolution~\cite{Peloso:2022ovc}. 

The evolution of the inflaton field, of the scale factor, and of the gauge modes is controlled by the last two equations in~(\ref{eom-phi-A}) and by eq.~(\ref{eq-Ak}). In Appendix \ref{app:codevar} we show how these equations are combined, and written in terms of rescaled variables, in our numerical integrations. 

Without loss of generality, we can assume $\phi' > 0$ in eq.~(\ref{eq-Ak}), so that the mode functions $A_+$ experience a large growth during inflation, while the modes $A_-$ remain close to their vacuum value, and we disregard them in the remainder of this work.\footnote{If instead $\phi' < 0$, the two polarizations interchange their role.} To express in a compact way the gauge field contributions to the last two equations of (\ref{eom-phi-A}), it is convenient to introduce the fields 
\begin{equation}
{\hat C}_i^E \equiv {\hat E}_i \left( \tau ,\, \vec{x} \right) 
\;\;,\;\; 
{\hat C}_i^B \equiv {\hat B}_i \left( \tau ,\, \vec{x} \right) \;, 
\end{equation} 
which evaluate to (using the property $i\,\vec{k}\times\vec{\epsilon} \,{}^{(\pm)} ( \vec{k} ) =\pm k\,\vec{\epsilon} \,{}^{(\pm)} ( \vec{k} )$ for the ``magnetic'' contribution) 
\begin{equation} 
{\hat C}_i^\alpha = \frac{1}{a^2} 
\int \frac{d^3k}{\left(2\pi \right)^{3/2}} \, {\rm e}^{i{\vec k\cdot \vec x}} \epsilon_i^{(+)}(\vec k) \left[ 
F_\alpha (\tau,\, k) \,{\hat a}_+ ( \vec k )  + 
F_\alpha^* (\tau,\, k) \,{\hat a}_+^\dagger ( -\vec k )
\right] \,, 
\label{C-EB}
\end{equation} 
with mode functions given by 
\begin{equation}
F_\alpha \left( \tau ,\, k \right) \equiv  \left\{ - A_+' \left( \tau ,\, k \right) ,\, k \, A_+ \left( \tau ,\, k \right) \right\} \;, 
\label{F-EB}
\end{equation} 
(with $\alpha = {\rm E} ,\, {\rm B}$). From this expressions one readily obtains 
\begin{equation}
\left\langle {\hat C}_i^\alpha \, {\hat C}_i^\beta \right\rangle_{\rm symm.} \equiv \left\langle \frac{{\hat C}_i^\alpha {\hat C}_i^\beta + {\hat C}_i^\beta {\hat C}_i^\alpha}{2} \right\rangle = \frac{1}{2 \pi^2 a^4} \int_0^\infty d k \, k^2 \, {\rm Re } \left[ F_\alpha \left( \tau ,\, k \right) F_\beta^* \left( \tau ,\, k \right) \right] \;, 
\label{CE-CB}
\end{equation} 
where ${\rm Re }$ denotes the real part (we note that the symmetrization is only required to evaluate the $\vec{E} \cdot \vec{B}$ product). 

In Appendix \ref{app:constxi} we evaluate the gauge mode functions and these correlators in the case of constant $\xi$.

\subsection{Equations for the GWs}
\label{subsec:eq-GW}

Decomposing the FLRW line element, perturbed by tensor modes, as 
\begin{equation}\label{lagr_tens}
d s^2 = a^2 \left( \tau \right) \left[ - d \tau^2 + \left( \delta_{ij} + \hat{h}_{ij} \left( \tau ,\, \vec{x} \right) \right) dx^i dx^j \right] \,, 
\end{equation}
where $\hat{h}_{ij}$ is transverse and traceless, we obtain, to quadratic order in ${\hat h}$, 
\begin{equation}
S_{\rm GW} = \int d^4 x \left[ \frac{M_p^2 a^2}{8} \left( \hat{h}_{ij}' \hat{h}_{ij}' - \hat{h}_{ij,k} \hat{h}_{ij,k} \right) - \frac{a^4}{2} \, \hat{h}_{ij}  \left(  \hat{E}_i \, \hat{E}_j +  \hat{B}_i \, \hat{B}_j \right) \right] \,. 
\end{equation} 

To obtain a canonically normalized field describing tensor modes in Fourier space we decompose 
\begin{equation}
\hat{h}_{ij} \left( \tau , \vec x \right) = \frac{2}{M_p \, a(\tau)} \int \frac{d^3 k}{\left( 2 \pi \right)^{3/2}} \, {\rm e}^{i \vec{k} \cdot \vec{x}} \sum_{\lambda = +,-} \Pi_{ij,\lambda}^* ( {\hat k} ) \, 
\hat{Q}_\lambda ( \tau,\, \vec{k} ) \,, 
\label{deco-hij}
\end{equation}
where the polarization operators, 
\begin{eqnarray} 
\Pi_{ij,\pm}^* ( {\hat k} ) \equiv \epsilon_i^{(\pm)} ( {\hat k} )  \epsilon_j^{(\pm)} ( {\hat k} ) \,,  
\label{pol_tensor}
\end{eqnarray}
are written in terms of the circular polarization operators introduced in eq.~(\ref{A-deco}). The equations of motion for $\hat{Q}_\lambda$ are then 
\begin{equation}
\left( \frac{\partial^2}{\partial \tau^2} + k^2 - \frac{a''}{a} \,  \right) \hat{Q}_\lambda ( \vec{k},\,\tau ) = - \frac{a^3}{M_p} \, \Pi_{ij,\lambda} ( {\hat k} )\int \frac{d^3x}{(2\pi)^{3/2}} \, {\rm e}^{-i \vec k \cdot \vec x}  \left[\hat{E}_i \, \hat{E}_j +  \hat{B}_i \, \hat{B}_j  \right] \equiv \hat{\cal S}_\lambda(\tau,\,\vec{k}) \,, 
\label{eom_Qlambda}
\end{equation} 
which are solved by the sum $\hat{Q}_\lambda = \hat{Q}_\lambda^{(v)} + \hat{Q}_\lambda^{(s)}$  of a homogeneous plus a sourced term. From the solution, we evaluate the power spectra   
\begin{eqnarray} 
P_\lambda \left( k \right) \delta^{(3)} ( \vec{k} + \vec{k}' ) &=& \frac{k^3}{2 \pi^2} \, 
\left\langle {\hat h}_\lambda ( \tau ,\, \vec{k} ) \, {\hat h}_\lambda ( \tau ,\, \vec{k}' ) \right\rangle \nonumber\\ 
&=& \frac{k^3}{2 \pi^2} \, \frac{4}{M_p^2 a^2} 
\left\langle {\hat Q}_\lambda ( \tau ,\, \vec{k} )  \, {\hat Q}_\lambda ( \tau ,\, \vec{k}' ) \right\rangle \;, 
\label{PS-lambda}
\end{eqnarray} 
which are the sum of the incoherent vacuum and sourced contributions. 

To compute the present contribution of the GWs one needs to account for the transfer functions for these modes from horizon re-entry to today~\cite{Boyle:2005se}. The GW modes of our interest re-enter the horizon during radiation domination. For these modes, one finds the present fractional energy density \cite{Caprini:2018mtu} 
\begin{equation}
\Omega_{\rm GW} h^2 \left( k \right) \equiv \frac{h^2}{\rho_c} \, \frac{d \rho_{\rm GW,k}}{d \ln k} = \frac{\Omega_R \, h^2}{24} \, \sum_\lambda P_\lambda \left( k \right) \;, 
\label{omGW}
\end{equation}
where $\rho_c$ is the critical energy density of the current universe, $h$ parametrizes the Hubble constant in units of $100 \, {\rm km} \, {\rm s}^{-1} \, {\rm Mpc}^{-1}$, while $\Omega_R \simeq 4.15 \times 10^{-5} / h^2$ is the current fractional density in photons and neutrinos, as if they were relativistic today (the convenience in this rescaling is dictated by the fact that the ratio between the energy density of GW and radiation is constant inside the horizon).

\subsection{Vacuum GWs} 
\label{subsec:vacuum-GW}

Disregarding slow-roll corrections, the homogeneous solution of eq.~(\ref{eom_Qlambda}), with mode functions normalized to the adiabatic vacuum at asymptotically early times, is 
\begin{eqnarray}
\hat{Q}^{(v)}_\lambda ( \vec{k} ) &=&  q_\lambda ( \tau,\, k ) \, \hat{a}_\lambda ( \vec{k} ) +  q_\lambda^* ( \tau,\, k ) \, \hat{a}_\lambda^\dagger ( - \vec{k} ) \,, \nonumber\\ 
q_\lambda ( \tau ,\, k ) &=& \frac{{\rm e}^{-i k \tau}}{\sqrt{2 k} } \left( 1 - \frac{i}{k \, \tau} \right)\,,
\label{sol_tensorvac}
\end{eqnarray} 
from which we write 
\begin{equation}
\left\langle {\hat Q}_\lambda^{(v)} ( \tau ,\, \vec{k} )  {\hat Q}_\lambda^{(v)} ( \tau ,\, \vec{k}' ) \right\rangle = \left\vert q_\lambda \left( \tau ,\, k \right) \right\vert^2 \delta^{(3)} \left( \vec{k} + \vec{k}' \right) \to \frac{H^2 a^2}{2 \,k^3} \, \delta^{(3)} \left( \vec{k} + \vec{k}' \right) \;, 
\end{equation} 
where the last expression is the super-horizon limit. Inserting this in (\ref{PS-lambda}), we obtain the well known expression for the vacuum power spectrum  
\begin{equation}
P_\lambda^{(v)} =  \frac{H^2}{\pi^2  M_p^2 } \;, 
\end{equation} 
which is independent of $\lambda$ (the vacuum is unpolarized). 

\subsection{Sourced GWs}
\label{subsec:sourced-GW}

The sourced solution of eq.~(\ref{eom_Qlambda}) is 
\begin{equation}
\hat{Q}^{(s)}_\lambda ( \tau ,\, \vec{k} ) = \int^\tau d \tau' \, G_k \left( \tau ,\, \tau' \right)\, \hat{\cal S}_\lambda(\tau',\,\vec{k})\,,
\label{Q1lambda_formal}
\end{equation} 
where the Green function satisfies 
\begin{equation}
\left( \frac{\partial^2}{\partial \tau^2} + k^2 - \frac{a''}{a} \right) G_k \left( \tau ,\, \tau' \right) = \delta^{(3)} \left( \tau - \tau' \right) \;,
\label{eq-Green-tau}
\end{equation} 
and, as we are interested in the retarded one, we write 
\begin{equation} 
G_k \left( \tau ,\, \tau' \right) \equiv  {\tilde G}_k \left( \tau ,\, \tau' \right) \, \Theta \left( \tau - \tau' \right) \;, 
\label{Green-theta}
\end{equation}
where $\Theta$ denotes the Heaviside step function and where ${\tilde G}_k$ is computed in Subsection \ref{subsec:Green}. The source $\hat{\cal S}_\lambda(\tau,\,\vec{k})$ is defined in eq.~(\ref{eom_Qlambda}). We rewrite it in terms of eq.~(\ref{C-EB}) and we evaluate the two-point correlator 
\begin{eqnarray} 
&& \!\!\!\!\!\!\!\!  \!\!\!\!\!\!\!\!  \!\!\!\!\!\!\!\!  \!\!\!\!\!\!\!\! 
\left\langle \hat{\cal S}_\lambda(\tau',\,\vec{k}) \, \hat{\cal S}_\lambda(\tau'',\,\vec{k}') \right\rangle = 
\frac{2 \, \delta^{(3)} \left( \vec{k} + \vec{k'} \right)}{a \left( \tau' \right) a \left( \tau'' \right)  M_p^2} 
\int \frac{d^3 p }{\left( 2 \pi \right)^3} 
\frac{1}{16} \left( 1 - \lambda \frac{p \cos \theta - k}{\sqrt{k^2-2 p \, k \cos \theta + p^2}} \right)^2 
\left( 1 + \lambda \cos \theta \right)^2 \nonumber\\ 
&& \quad\quad\quad\quad  \quad\quad\quad\quad  \quad\quad\quad\quad 
\left[ 
F_E (\tau',\, p) F_E (\tau',\, \vert \vec{k} - \vec{p} \vert ) + 
F_B (\tau',\, p) F_B (\tau',\, \vert \vec{k} - \vec{p} \vert )  
\right]   \nonumber\\ 
&& \quad\quad\quad\quad  \quad\quad\quad\quad  \quad\quad\quad\quad 
\left[
F_E (\tau'',\, p) F_E (\tau'',\, \vert \vec{k} - \vec{p} \vert ) + 
F_B (\tau'',\, p) F_B (\tau'',\, \vert \vec{k} - \vec{p} \vert )  
\right]^* \;,  \nonumber\\ 
\label{SS}
\end{eqnarray} 
where $\theta$ is the angle between the external momentum $\vec{k}$ and the integration momentum $\vec{p}$, and where the property 
\begin{equation}
\left\vert \vec{\epsilon}^{(\lambda)} \left( {\hat p} \right) \cdot \vec{\epsilon}^{(+)} (\vec q) \right\vert^2
= \left( \frac{1- \lambda \, {\hat p} \cdot {\hat q}}{2} \right)^2  \,,
\end{equation} 
has been used. Using eqs. (\ref{F-EB}), (\ref{PS-lambda}), and (\ref{Q1lambda_formal}), we then arrive to 
\begin{eqnarray} 
&& \!\!\!\!\!\!\!\! 
P_\lambda^{(s)} \left( k \right)  = \frac{1}{4 \pi^2 M_p^4} 
\int \frac{k \, d^3 p}{\left( 2 \pi \right)^3} 
\left( 1 - \lambda \frac{p \cos \theta - k}{\sqrt{k^2-2 p \, k \cos \theta + p^2}} \right)^2 
\left( 1 + \lambda \cos \theta \right)^2 \nonumber\\ 
&&  \!\!\!\!\!\!\!\! 
\times\left\vert \int^{k \tau} d \left( k \tau' \right) 
\frac{k \,{\tilde G}_k \left( \tau ,\, \tau' \right)}{a \left( \tau \right) a \left( \tau' \right)}
\left[ \frac{1}{k} \, A_+' (\tau',\, p) A_+' (\tau',\, \vert \vec{k} - \vec{p} \vert ) + 
\frac{p \,  \vert \vec{k} - \vec{p} \vert }{k}  \, A_+ (\tau',\, p) A_+ (\tau',\, \vert \vec{k} - \vec{p} \vert )  
\right] \right\vert^2 \,. \nonumber\\ 
\label{P-la-main}
\end{eqnarray} 
The rescalings in this expression are motivated by the fact that, in natural units, the functions $A$ and ${\tilde G}_k$ have, respectively, mass dimension $-1/2$ and $-1$, so that the second line of this expression is dimensionless. We then immediately see that the full expression (\ref{P-la-main}) is dimensionless. 

In Appendix \ref{app:codevar} we rewrite this expression in terms of the variables used in our numerical integrations. In Appendix \ref{app:constxi} we instead evaluate this expression in a de Sitter background and for constant $\xi$, showing that it leads to a scale invariant power spectrum outside the horizon, in agreement with the results in the literature.

\subsection{Green function for the sourced GWs}
\label{subsec:Green}

In this subsection we study the Green function of eq.~(\ref{eq-Green-tau}). In particular, we are interested in the retarded Green function, that is written in the form (\ref{Green-theta}), where the function ${\tilde G}_k$ needs to satisfy 
\begin{equation}
\left( \frac{\partial^2}{\partial \tau^2} + k^2 - \frac{a''}{a} \right) 
{\tilde G}_k \left( \tau ,\, \tau' \right) = 0 \;\;,\;\; 
{\tilde G}_k \left( \tau ,\, \tau \right) = 0 \;\;,\;\; 
\frac{d}{d \tau} {\tilde G}_k \left( \tau ,\, \tau' \right) \Big\vert_{\tau' = \tau} = 1 \;. 
\label{conditions-Green}
\end{equation} 

To obtain this function, we consider two linearly independent solutions $F_{1,2} \left( \tau ,\, k \right)$ of the associated homogeneous equation  
\begin{equation}
\left( \frac{\partial^2}{\partial \tau^2} + k^2 - \frac{a''}{a} \right) 
F_i \left( \tau  ,\, k \right) = 0 \;\;,\;\; i = 1 ,\, 2 \;. 
\label{eq-F-tau}
\end{equation}
In particular, for any given mode (namely, for any given value of $k$), the last term $\frac{a''}{a}$ is much smaller than $k^2$ at asymptotically early times (when the mode is deep inside the horizon), and we can choose for $F_1 \left( \tau \right)$ the solution that approaches the positive-frequency adiabatic vacuum at these early times 
\begin{equation}
\lim_{\tau \to - \infty} F_1 \left( \tau  ,\, k \right) = \frac{{\rm e}^{-i k \tau}}{\sqrt{2 k}} \;. 
\label{adibatic-in}
\end{equation}
We then choose $F_2 \left( \tau,\,k \right) = F_1^* \left( \tau  ,\, k \right)$, noting that $F_1$ and $F_1^*$ are indeed linearly independent. We note from the asymptotic form (\ref{adibatic-in}) that these two solutions satisfy the Wronskian condition 
\begin{equation}
F_1' \left( \tau  ,\, k \right) \, F_2 \left( \tau  ,\, k \right) - 
F_2' \left( \tau  ,\, k \right) \, F_1 \left( \tau  ,\, k \right) = - i \;. 
\label{wronskian}
\end{equation} 
Using eq.~(\ref{eq-F-tau}) it is immediate to see that the left hand side of this expression is constant, and therefore the two solutions $F_{1,2} \left( \tau \right)$ satisfy eq.~(\ref{wronskian}) at all times. Using this, we can then see that 
\begin{equation}
{\tilde G}_k \left( \tau ,\, \tau' \right) = \frac{F_1 \left( \tau  ,\, k \right) F_2 \left( \tau'  ,\, k \right) - F_2 \left( \tau  ,\, k \right) F_1 \left( \tau'  ,\, k \right)}{F_1' \left( \tau'  ,\, k \right) F_2 \left( \tau'  ,\, k \right) - F_2' \left( \tau'  ,\, k \right) F_1 \left( \tau'  ,\, k \right)}  = 2 \, {\rm Im } \left[ F_1^* \left( \tau  ,\, k \right) \, F_1 \left( \tau'  ,\, k \right) \right]\;, 
\end{equation} 
where ${\rm Im }$ denotes the imaginary part, satisfies all the requirements in (\ref{conditions-Green}). Therefore, combining the above results, the Green function of our interest is given by 
\begin{equation}
G_k \left( \tau ,\, \tau' \right) = {\tilde G}_k \left( \tau ,\, \tau' \right) \, 
\Theta \left( \tau - \tau' \right) \;\;,\;\; 
{\tilde G}_k \left( \tau ,\, \tau' \right) =  2 \, {\rm Im } \left[ F_1^* \left( \tau  ,\, k \right) \, F_1 \left( \tau'  ,\, k \right) \right]\;, 
\label{Green-tau}
\end{equation}
where $F_1 \left( \tau ,\, k \right)$ is the solution of eq.~(\ref{eq-F-tau}) subject to the initial condition (\ref{adibatic-in}). 

We stress that eq.~(\ref{Green-tau}) is the Green function for the sourced GW for any arbitrary evolution of the scale factor $a \left( \tau \right)$, and not simply for slow-roll inflation. One only requires a sufficiently prolonged accelerated expansion stage so that initially the modes were deep inside the horizon, and the initial condition~(\ref{adibatic-in}) is physically motivated. In particular, it can be used for our problem, where the inflaton and, consequently, the scale factor have a nontrivial evolution. 

In Appendix \ref{app:codevar} we rescale this Green function and rewrite it in terms of the variable used in our numerical integrations. Here we make two observations. Firstly, we show that eq.~(\ref{Green-tau}) reproduces the known Green function in a de Sitter background. In this case
\begin{equation}
a_{\rm dS} = - \frac{1}{H \tau} \;\;\; \Rightarrow \;\;\; 
F_{1,{\rm dS}} \left( \tau ,\, k \right) = \frac{1}{\sqrt{2 k}} \left( 1 - \frac{i}{k \tau} \right) \, {\rm e}^{-i k \tau} \;, 
\end{equation} 
and, from eq.~(\ref{Green-tau}), 
\begin{equation}
{\tilde G}_{k,{\rm dS}} \left( \tau ,\, \tau' \right) = \frac{\left( 1 + k^2 \tau \tau' \right) \sin \left( k \left( \tau - \tau' \right) \right) - k \left( \tau - \tau' \right) \cos \left( k \left( \tau - \tau' \right) \right)}{k^3 \tau \tau'} \;, 
\label{Green-dS}
\end{equation} 
in agreement with the known result. Secondly, we observe from this expression that the Green function greatly simplifies in the super-horizon limit 
\begin{equation}
\lim_{k \tau ,\, k \tau' \to 0} {\tilde G}_{k,{\rm dS}} \left( \tau ,\, \tau' \right) = \frac{\tau^3-\tau^{'3}}{3 \tau \tau'} \;. 
\label{Green-dS-super}
\end{equation} 

We note that this result is $k-$independent. We show now that this is the case for any evolution of the scale factor. The most general solution of eq.~(\ref{eq-F-tau}) in the super-horizon limit is 
\begin{equation}
\lim_{k \ll a \left( \tau \right) H \left( \tau \right)} F \left( \tau ,\, k  \right) = a \left( \tau, \, k \right) \left[ c_1 + c_2 \int_{\tau_*}^\tau \frac{d \tau'}{a^2 \left( \tau' \right)} \right] \;, 
\end{equation} 
where $c_1$ and $c_2$ are two integration constants, while $\tau_*$ denotes some reference time after which the super-horizon approximation is accurate. The integration constants generally depend on $k$ and on the evolution of the scale factor before the mode entered in the super-horizon regime. However, they drop from the combination 
\begin{eqnarray}
\lim_{k \ll a \left( \tau \right) H \left( \tau \right), \, a \left( \tau'\right) H \left( \tau'\right)} {\tilde G}_k \left( \tau ,\, \tau' \right) &=& 
\lim_{k \ll a \left( \tau \right) H \left( \tau \right), \, a \left( \tau'\right) H \left( \tau'\right)} \frac{{\rm Im } \left[ F \left( \tau ,\, k \right) F^* \left( \tau' ,\, k \right) \right]}{{\rm Im } \left[ F' \left( \tau' ,\, k \right) F^* \left( \tau' ,\, k \right) \right]} \nonumber\\
&=&  
a \left( \tau \right) a \left( \tau' \right) \int_{\tau'}^\tau \frac{d \tau''}{a^2 \left( \tau'' \right)} \;, 
\label{Green-tau-super}
\end{eqnarray}
which is indeed $k-$independent, and which, we stress, is valid beyond slow-roll inflation. This relation reproduces the expression (\ref{Green-dS-super}) when evaluated in the de Sitter background, and is confirmed by our numerical integrations that we discuss below. After having verified it, we use the relation (\ref{Green-tau-super}) in the super-horizon regime, to speed up the numerical integrations.

\section{The numerical scheme} 
\label{sec:numerics} 

Evolving the system of equations (\ref{code-eqs-phi-A}) and (\ref{electro-vevs}) numerically during inflation in the strong backreaction regime is a highly non-trivial task.  Several different approaches have been proposed in the literature to tackle this problem. The first such approach was pursued in references \cite{Cheng:2015oqa,Notari:2016npn,DallAgata:2019yrr} and consisted of solving the resulting integro-differential system of equations by discretizing the integral and reconstructing it at every moment in time using the trapezoid or parallelogram rule. The second approach developed in \cite{Sobol:2019xls,Gorbar:2021rlt} makes use of the gradient expansion formalism, which consists of evolving a system of differential equations of bilinear electromagnetic functions. The third approach, presented in \cite{Domcke:2020zez}, solves the system of integro-differential equations in an iterative manner starting from the solution using the analytic approximation for the backreaction as the first step of the iteration. Finally, a full lattice calculation was performed in \cite{Caravano:2022epk}, which not only takes into account the spatial gradient of the gauge field but also that of the inflaton. This simulation has been used to compute the power spectrum of scalar perturbations as opposed to gravitational waves, which is the subject of the present work. Moreover, they can cover a dynamical range of only a few e-folds.~\footnote{A smaller dynamical range is required to study the consequences of the gauge field instability for the early stages of reheating after inflation. Lattice studies of axion-gauge field systems focusing on the epoch of reheating include~\cite{Adshead:2015pva,Adshead:2016iae,Figueroa:2017qmv,Adshead:2018doq,Cuissa:2018oiw,Adshead:2019lbr,Adshead:2019igv,Figueroa:2021yhd}.}

In order to facilitate our study of the power spectrum of gravitational waves at interferometer scales, we set a number of goals that our numerical scheme ought to achieve. Firstly, we want to have sufficient resolution and efficiency to study a strong backreaction regime which may be relevant for a large portion of the observable $\sim 60$ e-folds of inflation. We also want our code to avoid iterative work so that a single run is sufficient to get the full result. In order to achieve this we employ the first method described above and implement a number of techniques that improve upon the previous work \cite{DallAgata:2019yrr}. We implement our algorithm in \textit{Mathematica}.

\subsection{Comoving momentum discretization and regularization of the backreaction integral} 
\label{sec:discretization} 

Our discretization scheme for the backreaction integral consists of defining a set of comoving momenta in the code variable $\tilde{k}$, defined in eq.~(\ref{cod-var}), labeled with index $i$, equally spaced in logarithmic space. Namely,
\begin{eqnarray}
    \tilde{k}_i=\tilde{k}_{\rm min}\left(\frac{\tilde{k}_{\rm max}}{\tilde{k}_{\rm min}}\right)^{\frac{i-1}{i_{\rm max}-1}}\;, 
    \label{sampling}
\end{eqnarray}
where $i$ is an integer ranging from $1$ to $i_{\rm max}$. The minimum and maximum momentum should be generally chosen in such a manner to allow for sufficient resolution of the strong backreaction regime. It is not hard to obtain an indication for these upper and lower momenta.  We define $\tilde{k}_{\rm thr}(N)$ to be the comoving momentum at the interface between stable and unstable modes by setting the dispersion relation in (\ref{eq-Ak}) equal to zero. Converting this value in code variables (see Appendix \ref{app:codevar}) we get
\begin{eqnarray}
    \tilde{k}_{\rm thr}(N)\equiv {\rm e}^N \tilde{H}  \frac{d \tilde{\phi}}{dN}\;, 
    \label{eq:kthr}
\end{eqnarray}
where the scale factor has been set to one at the start of the evolution. The lower bound on the momentum range is given by the request that all modes at the start of our numerical integration are deep in the stability regime, while the upper bound is approximately the mode that exits the horizon at the end of inflation. We target a typical model of high-scale inflation with about $60$ e-folds of evolution, although these considerations can be clearly modified for runs of different duration. In terms of code variables, this results in
\begin{eqnarray}
    \tilde{k}_{\rm min}\gg \tilde{k}_{\rm thr}(0) \;\;\;\; {\rm and} \;\;\;\;  \tilde{k}_{\rm max}\lesssim {\rm e}^{60} \tilde{H}(0)\;.
    \label{range-k}
\end{eqnarray}

After experimenting with the total number of modes we find that about $i_{\rm imax}\simeq 400$ is sufficient to obtain convergent results for at least the last 40 e-folds of inflation (which are those for which we have a relevant backreaction in the cases we studied) and, for concreteness, we use $\tilde{k}_{\rm min}= 10^{4} \;\tilde{k}_{\rm thr}(0)$ and $\tilde{k}_{\rm max}={\rm e}^{60} \tilde{H}(0)$. 

Additionally the integrand of the backreaction term $\left\langle \vec{E} \cdot \vec{B} \right\rangle$, given in code variables in eq.~(\ref{electro-vevs}), generally has to be regularized in the ultra-violet (UV). In principle such a regularization may seem unnecessary since the quantity $\frac{d}{d N} \left\vert {\bar A}_\lambda \right\vert^2$ formally vanishes deep in the stable regime. However, there is always some small non-zero numerical noise which is then amplified in the UV due to the presence of the $d \tilde{k} \;\tilde{k}^2$ factor, combined with the fact that the comoving momentum range spans many decades in $k$-space. For these reasons it is necessary to regularize the integral, which in practice consists of setting a hard momentum cutoff beyond which any contribution is neglected.~\footnote{The use of the cut-off is justified by the fact that, when the backreaction is relevant, there is a clear separation between the peak of the scales that are physically amplified and the vacuum modes, see Figure \ref{fig:backreaction-spectrum}.}  The function $\tilde{k}_{\rm reg}(N)$ denotes the time dependent momentum threshold beyond which the gauge modes should be disregarded in the backreaction term. Physically, at every instant in time this threshold should be the maximum comoving momentum that has ever become tachyonic at any point in time before that instance, or mathematically
\begin{eqnarray}
    \tilde{k}_{\rm reg}(N)\equiv{\rm MAX}\left[\tilde{k}_{\rm thr}(N')\right], \;\;\;\; {\rm where} \;\;\;\; N'<N\;.
    \label{eq:increasing-cutoff}
\end{eqnarray}
The algorithmical implementation of this is described in detail in Appendix \ref{app:numerics1}.

\subsection{Gauge mode equations of motion regularization} 
\label{sec:A-eom-regularization} 
%
Evolving the equation of motion of the gauge modes (last of equations~(\ref{code-eqs-phi-A})) and the auxiliary functions $\bar{{\cal F}}(N,\tilde{k})$ (first of (\ref{eqs-N-barF})), necessary for the numerical computation of the Green function using (\ref{code-N-PS}), is rather challenging due to the large hierarchy of terms involved on the right hand sides. More specifically, both equations contain terms proportional to $\tilde{k}/(\tilde{H}{\rm e}^N)$ which, for the highest momentum modes in our code, are initially many orders of magnitude greater than one.

To expand on the previous point, the comoving momenta employed in our integration span a range of approximately 27 decades, cf.~eq.~(\ref{range-k}), and hence it is inevitable that at early times the hierarchy between the Hubble horizon scale and the smallest wavelength~\footnote{Namely the ratio between the initial $a \, \tilde{H}$ and the greater ${\tilde k}_i$, which need to be part of our range, since they become dynamically relevant in the later stages of the evolution.} is so large that keeping track of the evolution of the modes becomes exponentially more “expensive” computationally. In order to tackle this issue we keep the modes fixed at their vacuum configuration until they come close to the threshold between stability and instability, whereupon they are allowed to evolve. Specifically, the momentum below which the mode is free to evolve should be as low as possible, because that makes the computation faster, while simultaneously much greater than the threshold momentum $\tilde{k}_{\rm thr}$, which signifies the onset of the instability, in order to satisfy the assumption that the modes are well into their vacuum configuration. After several trials, we identified a convenient cutoff value to be
\begin{eqnarray}
    \tilde{k}_{\rm vac}(N)\equiv 10^{5/2}\; \tilde{k}_{\rm reg}(N)\;.
    \label{eq:vac-cutoff}
\end{eqnarray}

In practice, a mode ${\tilde k}_i$ that is initially greater than $\tilde{k}_{\rm vac}(N)$ is not integrated. As soon as $\tilde{k}_{\rm vac}(N)$ increases above ${\tilde k}_i$, the gauge mode with this momentum is initialized according to the last two initial conditions in (\ref{initial1}), and it is evolved according to the last of eqs. (\ref{code-eqs-phi-A}). Only after also $\tilde{k}_{\rm reg} (N)$ becomes greater than  ${\tilde k}_i$, we include the contribution of this mode in the backreaction term $\left\langle \vec{E} \cdot \vec{B} \right\rangle$ and in the GWs source. This process introduces a phase difference between the various modes that is however phenomenologically irrelevant for the computation of gravitational waves since the phase difference is factorized and then cancelled away in the expression for the gravitational wave power spectrum. To conclude this subsection, we note that the particular method by which we keep the modes frozen in the vacuum configuration employs a “trick” that takes advantage of the fact that in the vacuum configuration the first derivative of the rescaled gauge mode mode functions $\bar{A}$ and auxiliary functions $\bar{\cal F}$ is formally zero. We outline how we employ this property to keep the modes frozen in Appendix \ref{app:numerics1}.

\subsection{Summary of the numerical scheme}
\label{sec:numerics-summary}

The techniques outlined above, combined with several further simplifications and optimizations that are \textit{Mathematica} specific, described in details in Appendix \ref{app:numerics2}, result in a code that can evolve the strong backreaction regime which may last for a significant fraction of the total duration of inflation in a memory efficient manner. 

The code takes typically thirty minutes to run on a regular desktop computer and produces the background functions as well as the gauge modes and auxiliary functions $\bar{{\cal F}}(N,\tilde{k})$ as a function of time for each momentum $\tilde{k}_i$. These functions are then used in (\ref{code-N-PS}) combined with the Green function calculated (\ref{Green-N-formalsol}) to compute the power spectrum of gravitational waves. The power spectrum computation takes approximately 3.5 minutes per $\tilde{k}_i$,  which implies that the total power spectrum for the entirety of the $i_{\rm max}=400$ modes takes approximately $\sim 23$ hours to compute. For the example outlined in the next section we only compute the power spectrum for $250$ of the total modes, excluding those generated while $\xi$ was too small to give a visible GW production. For some further comments on the \textit{Mathematica} implementation see Appendix \ref{app:numerics2}.

\section{Results} 
\label{sec:results}

In this section we present the results of the code described above for a specific inflaton potential. We do not aim to be exhaustive in the study of a landscape of potentials, but we rather want to demonstrate how a simple potential with certain qualitative features may allow the system to enter the strong backreaction regime for some period during inflation, and produce characteristic signatures in the power spectrum of gravitational waves at scales relevant for pulsar timing arrays, astrometry, and space-based interferometers.


Qualitativaly, the potential is characterized by a very flat part at CMB scales, with a small and negative curvature which is consistent with CMB observations, followed by a steeper part during which the speed of the axion and, consequently, the particle production are enhanced. Subsequently, the potential flattens out and the production becomes inefficient, so to avoid PBH overproduction~\cite{Linde:2012bt,Garcia-Bellido:2016dkw}, GW oveproduction at reheating~\cite{Adshead:2019lbr,Adshead:2019igv} and the current limit on the SGWB at ground-based interferometers \cite{KAGRA:2021kbb}.

The exact shape of the potential beyond this point and a detailed study of the end of inflation is beyond the scope of this work. Our goal is to study in detail a period of inflation during which the system enters the strong backreaction regime. However, for concreteness, we are choosing a potential that is consistent with observations at CMB scales. 

\begin{figure}
    \centering
    \includegraphics[width=0.66\linewidth]{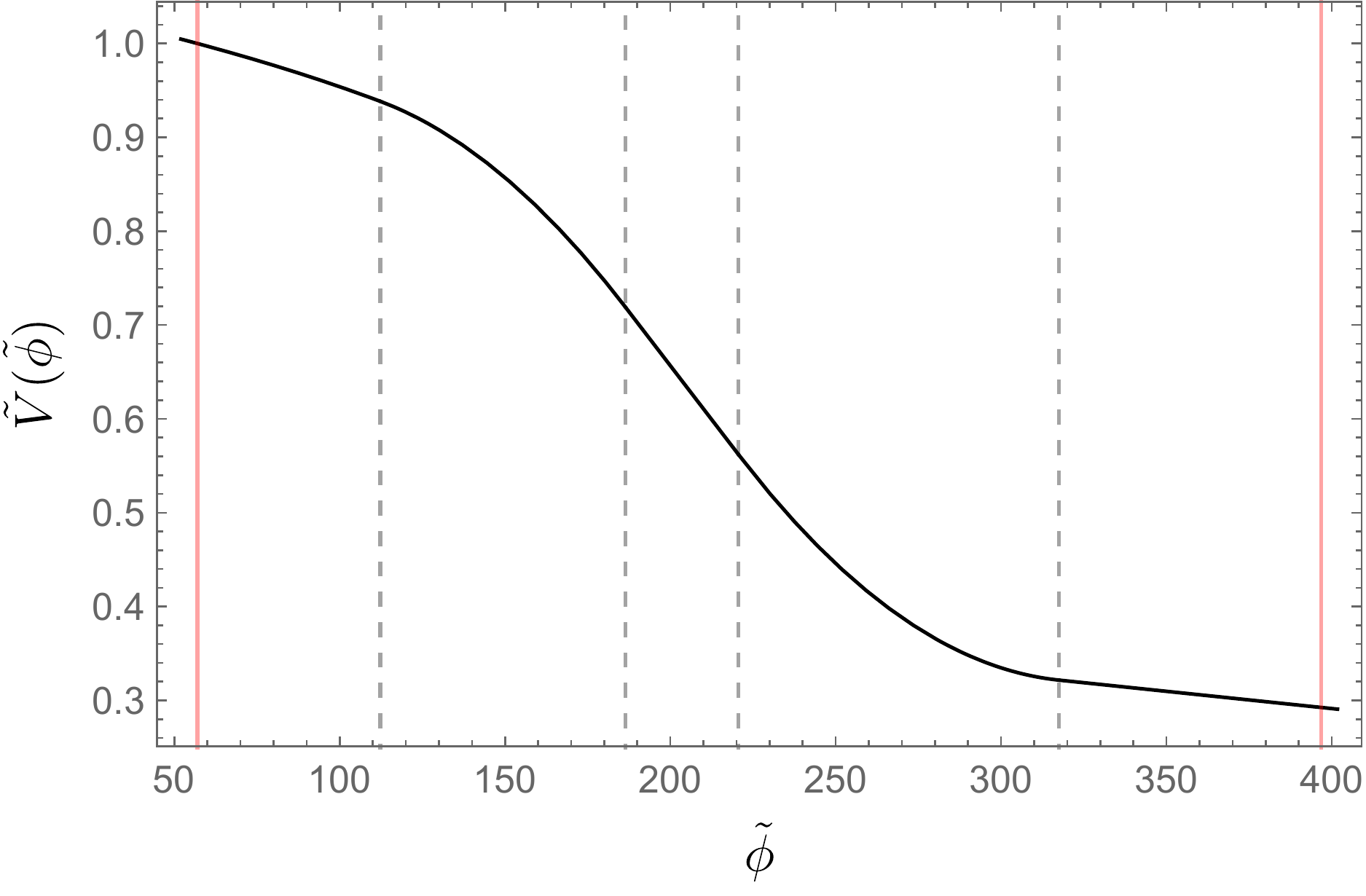}
    \caption{Plot of the five branch potential defined in (\ref{eq:potential}). The red vertical lines denote the field value sixty e-folds before the end of inflation, and at the end of inflation respectively. The black dashed lines denote the transition points between the various branches. The parameters in this example are $\tilde{\phi}_0=57,\,\tilde{\phi}_1=112,\,\tilde{\phi}_2=186,\,\tilde{\phi}_3=221,\,\tilde{\phi}_4=317,\, c_1= 9.48\cdot 10^{-4},\, c_2=2.39\cdot 10^{-6},\, c_3=9.05\cdot 10^{-9},\, c_4 = 3.97\cdot 10^{-11}$ and the two straight line slopes are $c_5=4.95\cdot 10^{-3}$ and $c_6=3.68 \cdot 10^{-4}$, respectively.}
    \label{fig:potential}
\end{figure}

Specifically, our potential consists of an initial quartic hilltop part, followed by two straight lines that are smoothly connected using quadratic expressions. This results in five branches
\begin{equation}
\tilde{V}(\tilde{\phi})= \left\{
\begin{array}{ll}
      \left[1-c_1 \left(\tilde{\phi}-\tilde{\phi}_0\right)-c_2\left(\tilde{\phi}-\tilde{\phi}_0\right)^2-c_3\left(\tilde{\phi}-\tilde{\phi}_0\right)^3-c_4 \left(\tilde{\phi}-\tilde{\phi}_0\right)^4\right] & \tilde{\phi}_0\leq \tilde{\phi} \leq \tilde{\phi}_1 \\
      p_1 \,\tilde{\phi}^2+p_2 \,\tilde{\phi}+p_3 & \tilde{\phi}_1 \leq \tilde{\phi} \leq \tilde{\phi}_2 \\
      c_5 \,\tilde{\phi}+p_4 & \tilde{\phi}_2 \leq \tilde{\phi} \leq \tilde{\phi}_3 \\
      p_5 \,\tilde{\phi}^2+p_6 \,\tilde{\phi}+p_7 & \tilde{\phi}_3 \leq \tilde{\phi} \leq \tilde{\phi}_4\\
      c_6 \,\tilde{\phi}+ p_8 & \tilde{\phi}_4 \leq \tilde{\phi}\;. \\
\end{array} 
\label{eq:potential}
\right. 
\end{equation}
Parameters $c_1-c_4$ are fixed so that at CMB scales the backreaction is negligible and the predictions are compatible with the Planck constraints on inflation. In turn, $c_5$ and $c_6$ are the slopes of the straight lines which control the amount of particle production at intermediate and late times, respectively. The remaining parameters $p_1-p_8$ are then fixed in terms of the previous ones to ensure continuity of the potential and its derivative at the interfaces of the five branches. We plot the potential in Figure~\ref{fig:potential}, listing the potential parameters in the caption.
The potential listed above, combined with the axion-gauge coupling strength $1/f=57/M_p$, define the parameters of our model example.

\begin{figure}
    \centering
    \includegraphics[width=0.45\linewidth]{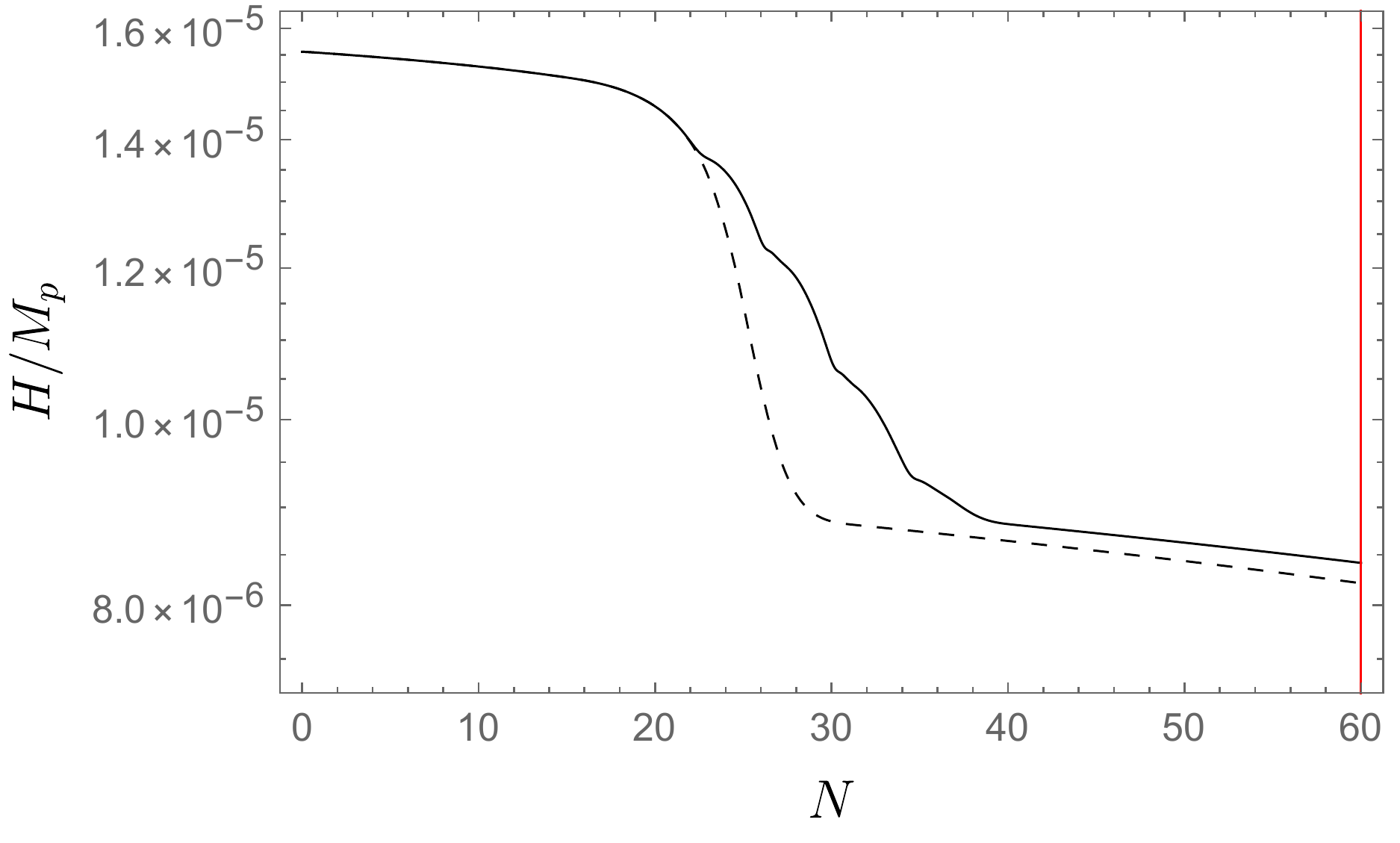}
    \includegraphics[width=0.45\linewidth]{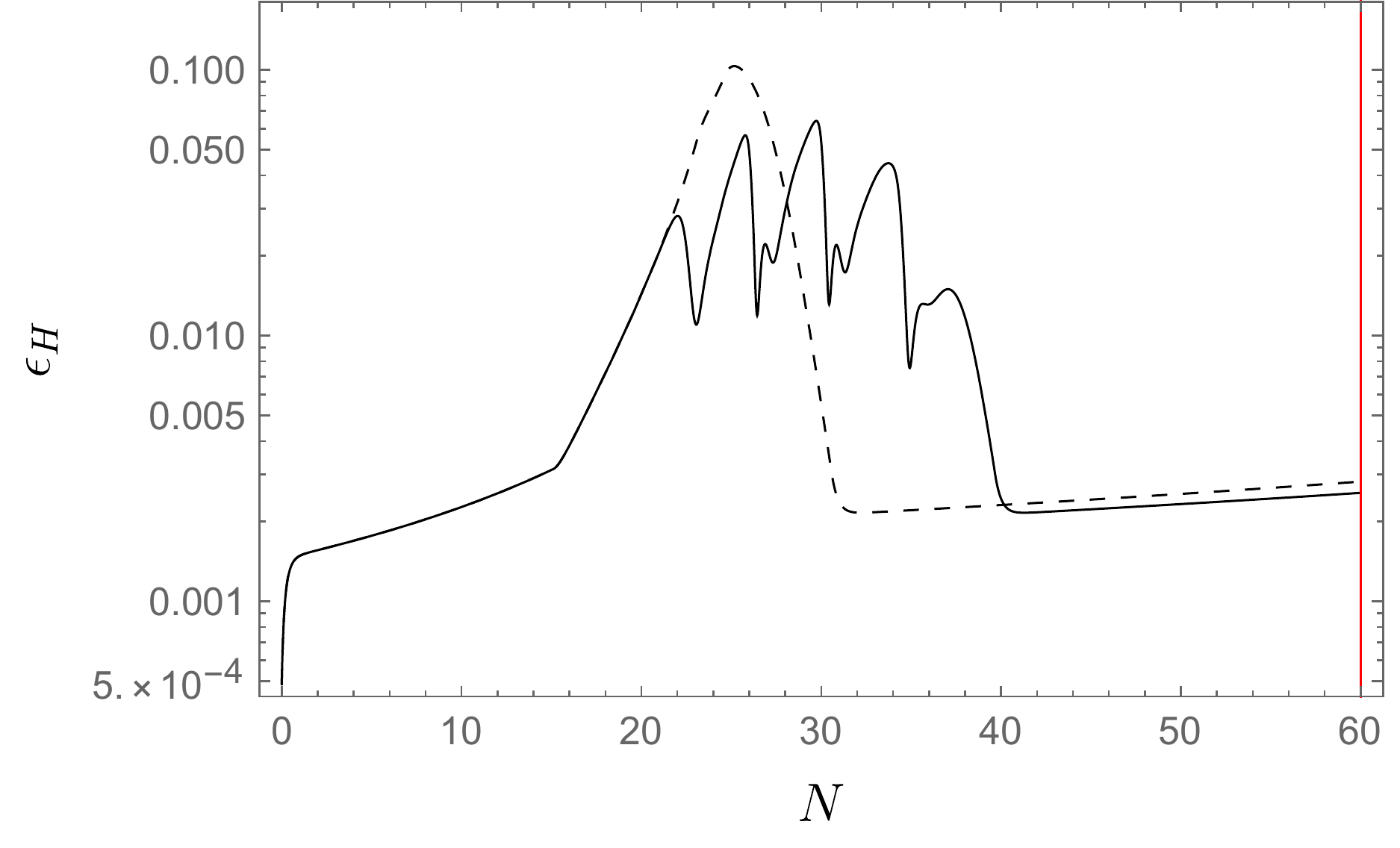}
    \includegraphics[width=0.66\linewidth]{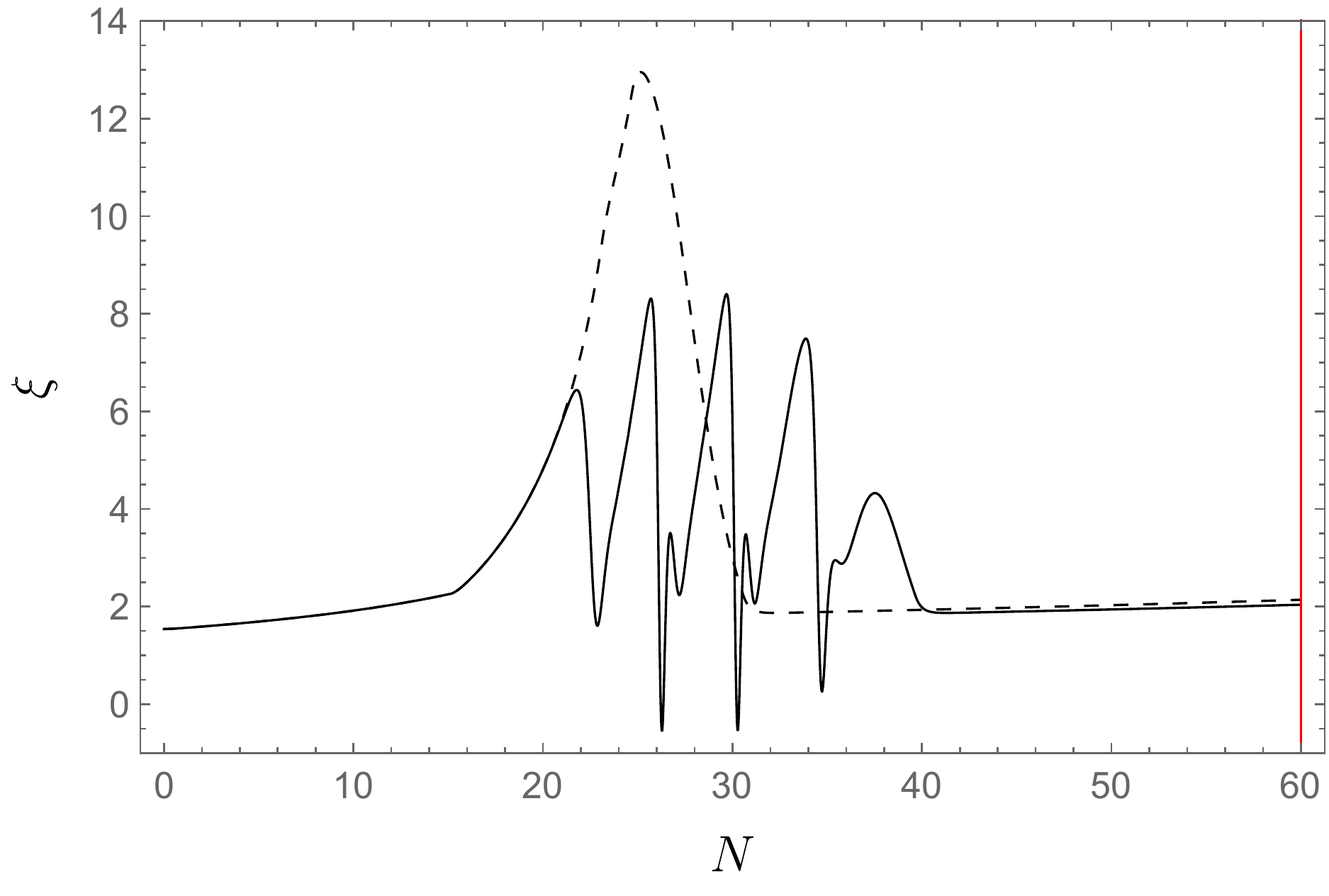}
    \caption{The two upper panels display, respectively, the evolution of the Hubble rate $H$ and the Hubble slow-roll parameter $\epsilon_H$ for the model example described in the main text. The bottom panel displays the evolution of the parameter $\xi \propto \dot{\phi} / H$ controlling the gauge field amplification. In all panels, the black solid lines take properly into account the backreaction of the produced gauge fields, while the black dashed lines show the evolution that would take place in the same potential if the backreaction were incorrectly disregarded. The evolution is shown as a function of the number of e-folds $N \equiv \ln a$, where the scale factor $a$ is normalized to $1$ at the start.}
    \label{fig:example-run}
\end{figure}

For these parameters, at CMB scales the system is in the low backreaction regime with CMB scale predictions $n_s=0.96$ and $r=0.023$ which are consistent with the latest constraints at $2\sigma$ \cite{BICEP:2021xfz}. Later in the evolution, when the slope of the potential becomes steeper, the system enters the strong backreaction regime. This regime is characterized by a nontrivial evolution of the inflaton speed that we have discussed in the Introduction, and that is visible through the evolution of $\xi \left( t \right)$ shown in the third panel of Figure \ref{fig:example-run}. In the first two panels of that figure we show instead the evolution of the Hubble rate $H$ and of the slow-roll parameter $\epsilon_H \equiv -\frac{\dot{H}}{H^2}$, confirming that this intermediate stage is not characterized by standard slow-roll evolution. 

\begin{figure}
    \centering
    \includegraphics[width=0.45\linewidth]{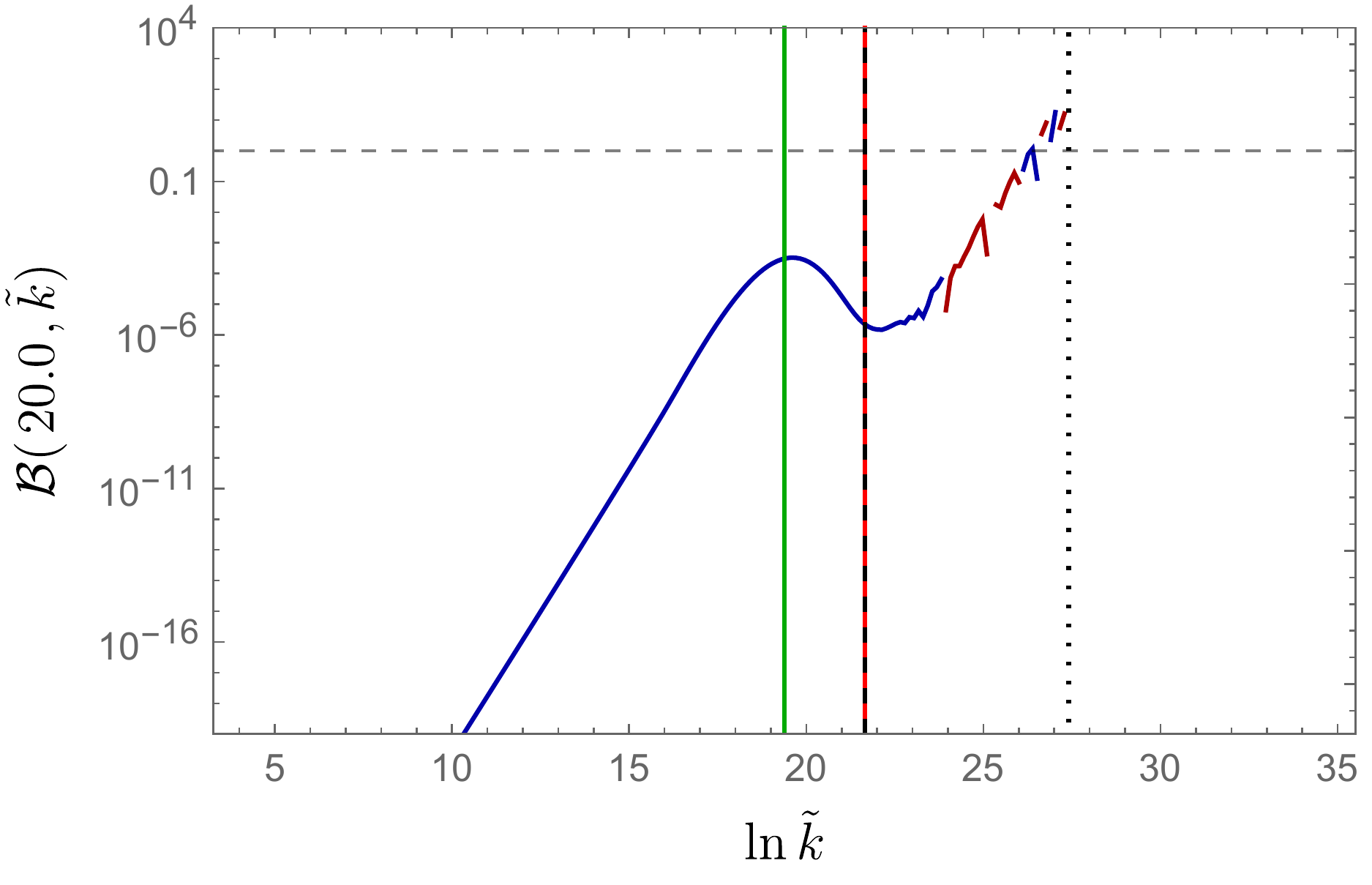}
    \includegraphics[width=0.45\linewidth]{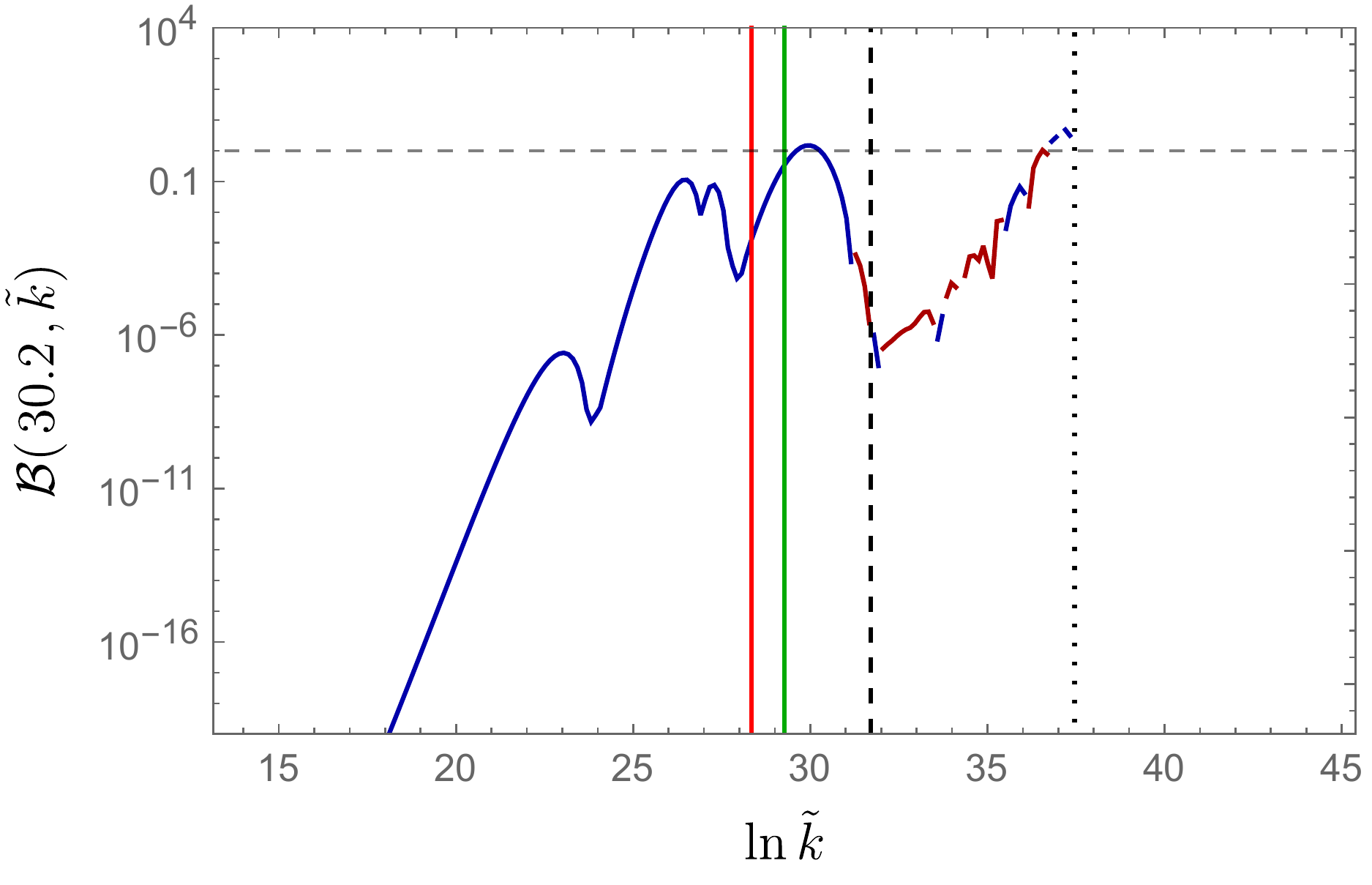}
    \includegraphics[width=0.45\linewidth]{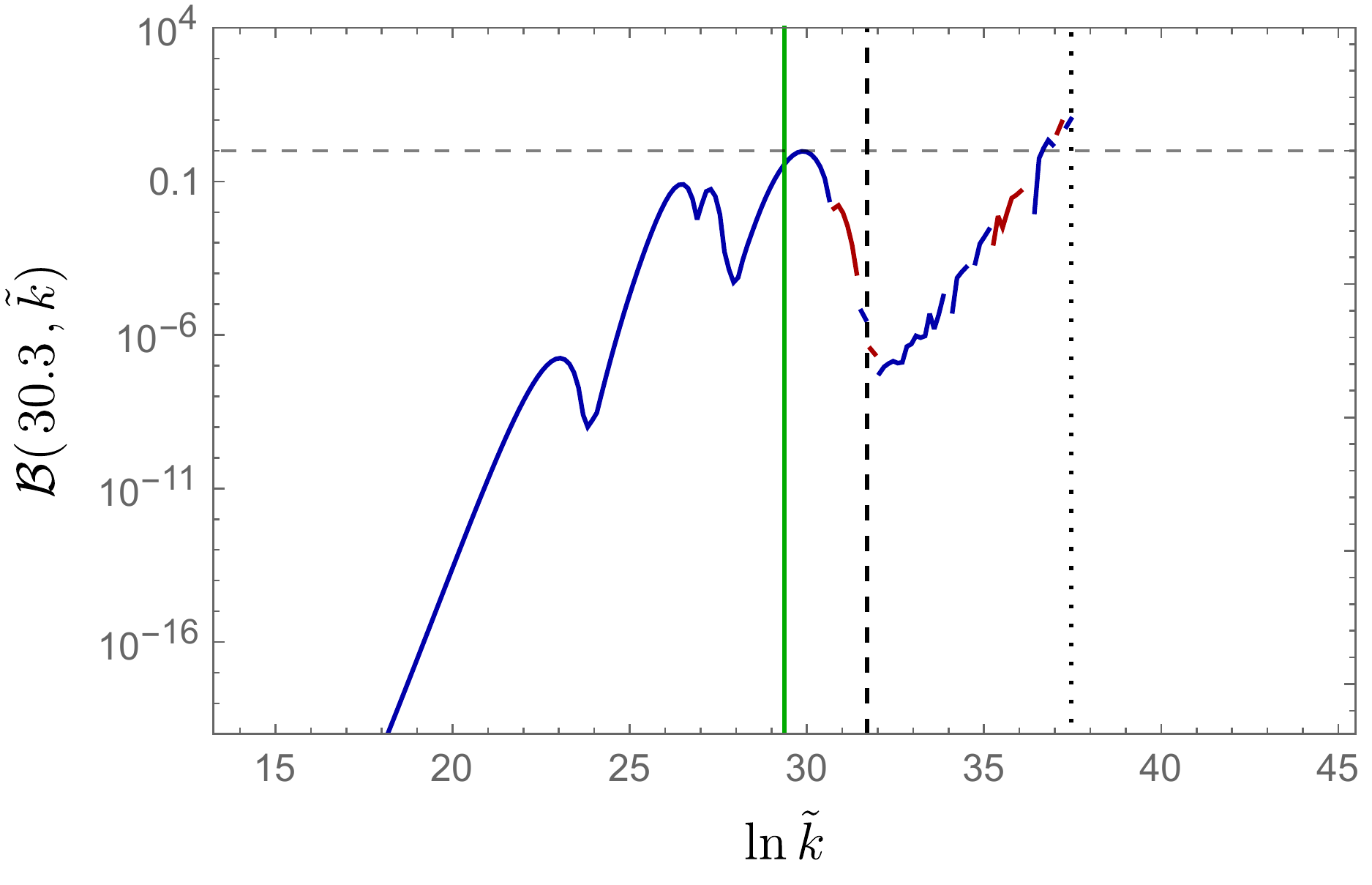}
    \includegraphics[width=0.45\linewidth]{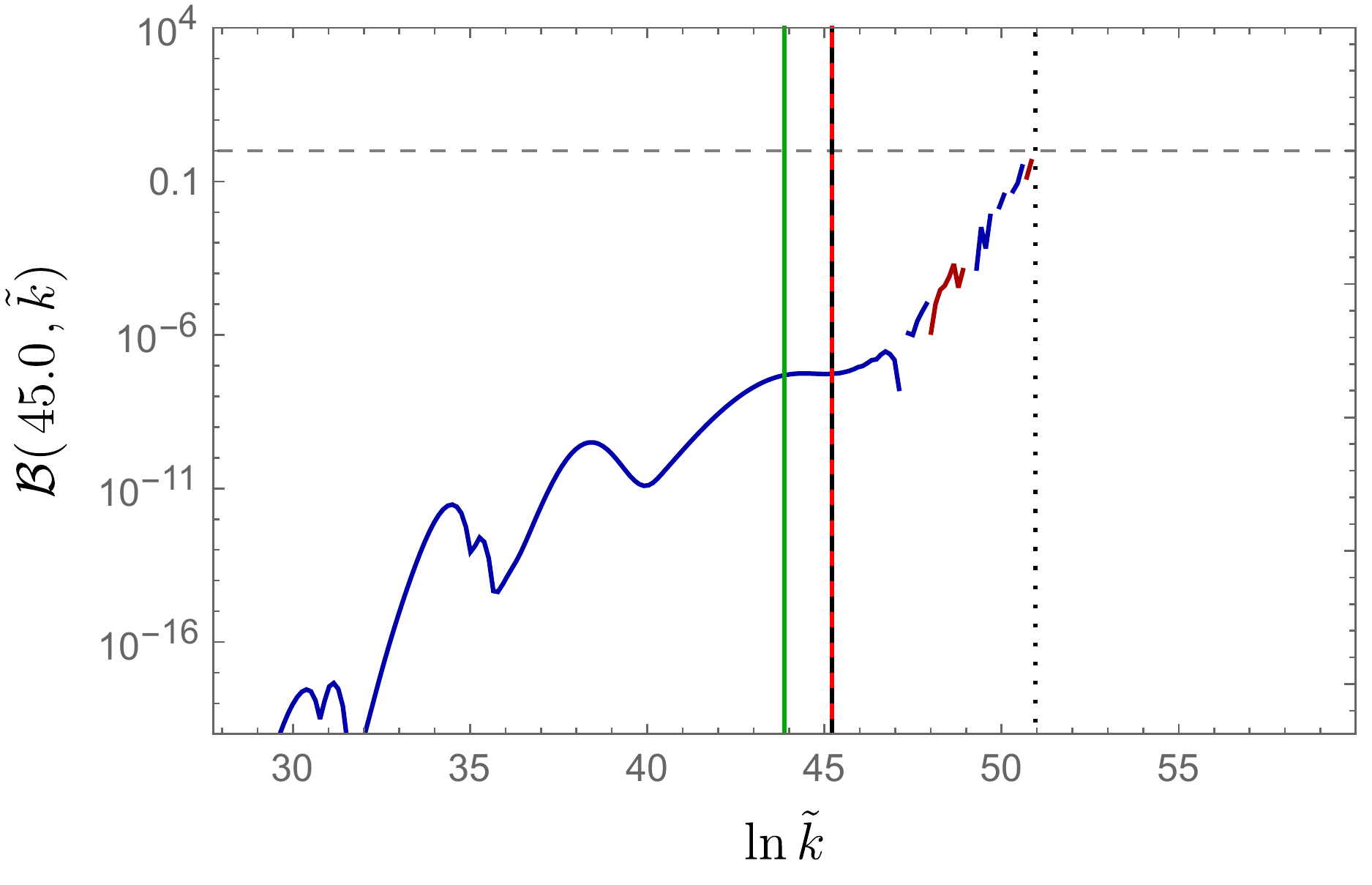}
    \caption{Plot of the backreaction spectrum defined in eq.~(\ref{eq:backreaction-spectrum}). Different panels correspond to different times, indicated by the value of $N$ on the vertical scale. For each panel: the blue (red) line corresponds to a positive (negative) contribution of a given mode to the backreaction at the end of inflation; the grey dashed horizontal line is the reference point that the backreaction spectrum must reach in order to be important in the evolution of the inflaton and is equal to one; The green line indicates the mode that crosses the horizon at that given moment; the solid red vertical line is the threshold between stability and instability $\tilde{k}_{\rm thr}$ given by eq.~(\ref{eq:kthr}); the black dashed vertical line is the backreaction cutoff $\tilde{k}_{\rm reg}$ defined in eq.~(\ref{eq:kreg}); the dotted black vertical line is the gauge mode vacuum evolution cutoff $\tilde{k}_{\rm vac}$ defined in eq.~(\ref{eq:vac-cutoff}).}
    \label{fig:backreaction-spectrum}
\end{figure}

This nontrivial evolution is due to the backreaction of the produced quanta. The backreaction term is shown in Figure \ref{fig:backreaction-spectrum} for a few indicative moments during inflation. The chronological orders of the four panels is top-left $\to$ top-right $\to$ bottom-left $\to$ bottom-right, and, for brevity, we will denote the four panels as ``first'', ``second'', ``third'', and ``fourth'' in this discussion, following the chronological order. In each panel we show the $k-$dependent integrand of the backreaction term, normalized with respect to the slope of the potential
\begin{eqnarray}
   \frac{\left\langle\vec{E}\cdot\vec{B}\right\rangle}{f\; V'(\phi)}\equiv \int {\rm d}\ln\tilde{k}\;{\cal B}(N,\tilde{k})\;,
   \label{eq:backreaction-spectrum}
\end{eqnarray}
so that when a significant portion of the spectrum ${\cal B}(N,\tilde{k})$ reaches the critical value of one, then also its integral is of order one, and we know that the system enters the strong backreaction regime. 

The first panel shows an early moment in time while the backreaction is negligible. The evolution of the backreaction spectrum is relatively straightforward during the stage of negligible backreaction, with the term dominated by the modes that have become unstable in the moments immediately before the one shown, and that are therefore close to the cut-off. The outmost left vertical solid (green) line corresponds to the horizon scale, ${\tilde k} = {\rm e}^N {\tilde H} \left( N \right)$, while the other vertical solid (red) line, ${\tilde k} = {\tilde k}_{\rm thr} \left( N \right)$, separates the unstable from the stable modes. These lines monotonically move to the right in this stage, analogously to the dashed vertical line ${\tilde k} = {\tilde k}_{\rm reg} \left( N \right)$, that indicates the upper limit of the modes included in the backreaction. This dashed line is defined as the greatest value ever assumed by the second solid (red) line, and therefore the two lines coincide as long as the second solid (red) line is moving monotonically to the right. We recall that modes between the dashed and the dotted vertical line, ${\tilde k} = {\tilde k}_{\rm vac} \left( N \right)$, are evolved by the code, but are not included in the backreaction, as these are still vacuum modes that need to be renormalized away. Finally, the gray, horizontal, dashed line is a visual reference point that denotes the value one in the vertical axis, to indicate when backreaction becomes important. 

The second and third panel are both taken at $N \simeq 30$. The backreaction term is now dominant, as shown by the fact that the backreaction spectra have reached the horizontal dashed line. We see from Figure \ref{fig:example-run} that at this moment the inflaton is experiencing a maximum of its speed for the third time. These three times have created three peaks in the gauge field spectrum, that in turn create the three peaks in the backreaction term that are visible in the two panels. We notice from the second panel that the backreaction has started to decrease the inflaton speed from this third maximum. This is testified by the fact that the vertical red line (corresponding to the value of $\xi \propto \dot{\phi}$ at the moment shown) has moved to the left of the vertical dashed line (corresponding to the maximum value tht $\xi \propto \dot{\phi}$ has ever attained up to that moment).

The third panel corresponds to a moment $\Delta N=0.1$ subsequent to the second one. The backreaction spectrum is nearly unchanged, but the red line has disappeared from the figure. This is due to the fact that the backreaction actually causes the inflaton speed to momentarily become negative for a very narrow interval of times around the one shown. The backreaction term can indeed dominate and overwhelm the equation of motion of the inflaton, practically eliminating its kinetic energy and in certain cases reversing its motion for brief instances (this has previously been noticed by the numerical analyses \cite{Cheng:2015oqa,DallAgata:2019yrr,Domcke:2020zez}). 

Lastly, the fourth panel corresponds to a late time in which the backreaction has again become negligible. The backreaction spectrum exhibits a series of peaks corresponding to number of times in which $\dot{\phi}$ reached a maximum during its evolution. This is again a stage of standard slow roll inflation, with $\xi \propto \frac{\dot{\phi}}{H}$ monotonically increasing, as witnessed by the superposition of the two red and dashed vertical lines. 

We note that the different panels show a different range of momenta. We also note that, at any fixed comoving momentum, the backreaction term decreases at late times (once the gauge field production at that scale has ceased) due to redshift.

\begin{figure}
    \centering
    \includegraphics[width=0.95\linewidth]{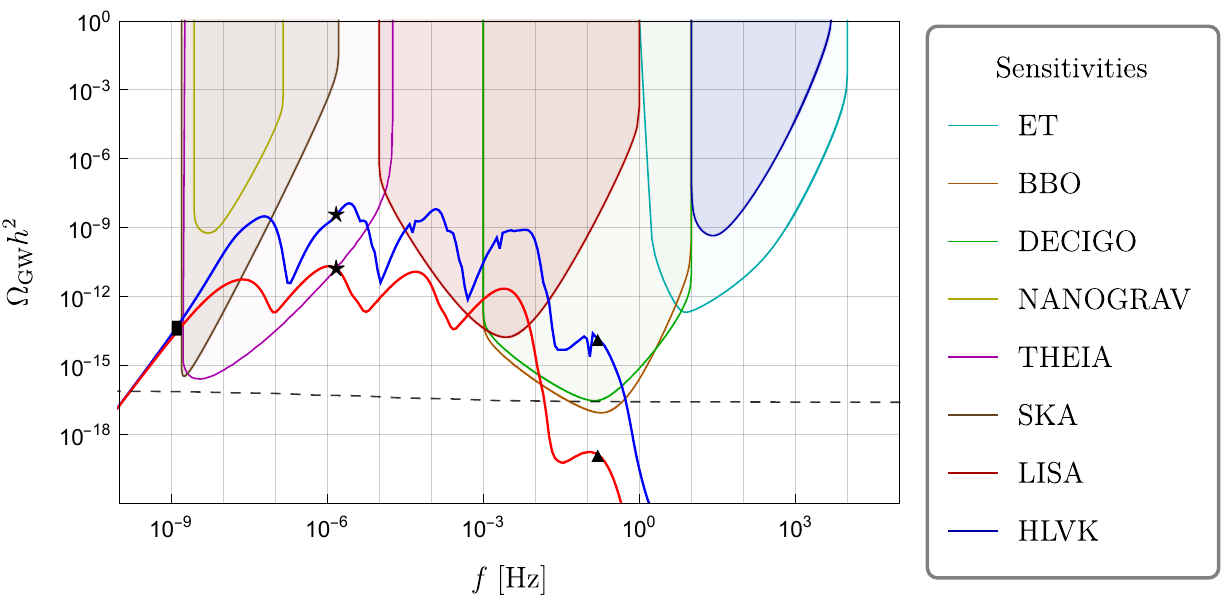}
    \caption{The sourced gravitational wave power spectrum of the \textit{Left} (blue line) and \textit{Right} (red line) polarization as well as the vacuum (back dashed line) superimposed with the power-law-integrated sensitivity curves corresponding to various current and future experiments taken from \cite{Schmitz:2020syl} and \cite{Garcia-Bellido:2021zgu} in the case of THEIA. The parameters chosen correspond to the coupling strength $1/f=57/M_p$. The squares, stars and triangles are points for which we plot the integrand of the power spectrum defined in (\ref{code-N-PS}) in Appendix \ref{app:parity}.}
    \label{fig:final-spectrum}
\end{figure}

Finally, let us discuss the SGWB produced in the model. The production is shown in Figure~\ref{fig:final-spectrum} in terms of the SGWB fractional spectral energy density (\ref{omGW}), with the blue and red curves corresponding to the two different GW polarizations and the black dashed curve corresponding to the vacuum contribution.

We see from the figure that the oscillatory features in the particle production parameter $\xi$ are also inherited by the sourced GWs spectrum, which also oscillates around an average value. The frequency of these oscillations matches the corresponding periodicity of the particle production parameter as a function of e-folds. For our choice of parameters the oscillations of the power spectrum cover a wide range of experiments from PTA, to astrometry, to space-based interferometers. The exact positioning of the peaks and the width of the overall signal depends on the exact shape of the potential and one can imagine that any potential whose slope varies in a qualitatively similar way to our example will necessarily imply the emergence of a power spectrum that shares the features observed in Figure \ref{fig:final-spectrum}. We see that, for essentially all the scales in which the sourced GWs are significant, one circular GW polarization is produced with  much greater amplitude than the other one. Although this is a well known result of this mechanism \cite{Sorbo:2011rz}, a novel effect that appears from our results is that the level of parity violation varies widely depending on the scales under consideration. Just to make an example, for the three points marked with, respectively, squares, stars and triangles in Figure \ref{fig:final-spectrum}, the chirality parameter 
\begin{eqnarray}
    \Delta \chi \equiv \frac{{\cal P}_{h,+}-{\cal P}_{h,-}}{{\cal P}_{h,+}+{\cal P}_{h,-}}\;,
    \label{eq:chirality}
\end{eqnarray}
is computed to be, respectively, $\Delta \chi = 0.25,\;\;0.992,\;\;0.99998$.

These features can be understood with a series of considerations. The most immediate one is the direct relation between the instantaneous value assumed by $\xi$ at any given time during inflation and the modes that were produced at that time. Consider two well separated times $t_1$ and $t_2$ and assume that $\xi$ takes the two different values $\xi_1$ and $\xi_2$ at those two times. For the moment, also assume for simplicity that $\xi$ is constant for some time around both $t_1$ and $t_2$. In this case, GWs of frequency $f_1$ (resp., $f_2$) that leave the horizons at times near $t_1$ (resp., $t_2$) are mostly sourced by gauge modes that also leave the horizon at that time, which have an amplitude controlled by $\xi_1$ (resp., $\xi_2$). Therefore, if $\xi_2 > \xi_1$, then $\Omega_{\rm GW} \left( f_2 \right) > \Omega_{\rm GW} \left( f_1 \right)$.

This effect is clearly seen in our results. There is however a further effect, related to the variation of $\xi$ within the few e-folds in which a given GW mode is mostly sourced. A GW mode of momentum $\vec{k}$ is sourced by two gauge modes, of momenta $\vec{p}$ and $\vec{q}$ that satisfy $\vec{p}+\vec{q} = \vec{k}$. If $\xi$ is rapidly growing while the scales $k$ leave the horizon, gauge modes of momenta greater than $k$ have a significantly greater amplitude than those of momenta of order $k$, and dominate the production of the GW modes of momentum $k$ (with the two vectors $\vec{p}$ and $\vec{q}$ being nearly anti-aligned). On the other hand, if $\xi$ is nearly constant or decreasing, the GW production is dominated by gauge modes of momenta close to or smaller than $k$. This can be seen in the shape of the integrand controlling the GW production, which we show and discuss in Appendix \ref{app:parity}. 

The integrands shown in that appendix also allow us to understand why the GW polarization visible in Figure \ref{fig:final-spectrum} is strongly scale-dependent. The GW modes with frequency marked with squares in the figure presenting nearly identical amplitude for the two polarization, are produced in a phase of fast growth of $\xi$. On the contrary, a significant polarization is obtained at scales marked with  stars and triangles, which were produced on a stage in which $\xi$ was, respectively, nearly constant (in an average sense, across a range of a few e-folds) and decreasing. As we just discussed, in the first case, the production is due to high-momentum, nearly head-on, gauge modes. These two gauge modes have nearly opposite helicity, and therefore the in-state has nearly vanishing total spin, producing a nearly equal amount of the two GW polarizations. In the other two cases, the production is dominated by gauge modes with momentum $\leq k$, resulting in a polarization-dependent GW production.

\subsection{Animation of the results}

The discussion of the previous subsection enlightens the interplay between the backreaction term and the evolution of the inflaton field. To further visualize this relation, and to get a more intuitive understanding of the system, we have prepared an animation in which the dynamical evolution of the backreaction spectrum is shown alongside that of the inflaton field and of the sourced GWs. Instead of the physical backreaction defined in (\ref{eq:backreaction-spectrum}) we plot the same quantity multiplied by ${\rm e}^{4N}/{\rm e}^{4\times 60}$, which factors out the dilution of the integrand of the backreaction due to the expansion of the universe, and which normalizes the overall multiplication factor to be equal to one at the end of inflation. This makes the visualization easier in video format, allowing us to keep the horizontal and vertical axis range fixed. We call this the comoving backreaction ${\cal B}_{\rm com}(N,\tilde{k})$ and the reference point that the backreaction line has to reach to be relevant for the equation of motion of the inflaton is now varying with time, and it is equal to ${\rm e}^{4N}/{\rm e}^{4\times 60}$ (gray horizontal dashed line).

We also plot the spectrum of gravitational waves in real time, once per e-fold, as it is being produced. We do this by computing the power spectrum starting from the beginning of inflation and then for each subsequent e-fold we add only the contribution of the intervening e-fold in the integral over the number of e-folds. This allows us to re-use the previous e-folds without having to compute the power from the start of inflation at every moment in time. This technique is possible due to the properties of the Green function which allows us to fix the time dependence of the Green function at $N=60$ inside the integral and extract the factor ${\rm e}^{N/60}$ outside the time integration. This is a good approximation as long as $N>N'$ which is true for our computation. We also verified that this incremental process yields the same final power spectrum as it does computing the power spectrum at the end of inflation in a single step. 

The video is available at \href{https://youtu.be/xRmiEhZEiRA}{this link}, where the various panels display the evolution of the inflaton along its potential (top left), the particle production parameter (top right), the backreaction spectrum (bottom left), and the gravitational wave power spectrum (bottom right) in “real time”. It consists of $50$ frames per second with each second corresponding to a single e-fold of evolution. 

\section{Conclusions} 
\label{sec:conclusions} 

An appealing scenario for early universe inflation is that in which an axion plays the role of the inflaton. Such a scenario, motivated by considerations of technical naturalness~\cite{Freese:1990rb}, introduces new dynamics through the axionic coupling of the inflaton to gauge fields. This coupling strongly amplifies one helicity of the gauge fields, making the overall dynamics of the system very non-trivial, with issues like backreaction that become important. Moreover, since the amplified gauge fields also naturally source gravitational waves, they provide a new way of generating a stochastic background of GW beyond that associated with vacuum fluctuations, with a greater spectrum at smaller scales than the CMB ones, that can be detected by a variety of GWs observatories. 

It has been recently realized that the effects of backreaction of the gauge fields on the inflaton-axion induces a {\em qualitative} change in the dynamics of the system. In this regime, the motion of the inflaton field and the gauge field production do not reach a steady state evolution, where the friction on the inflaton motion through the energy dissipated in the gauge fields perfectly balance each other at every time. Rather, the system performs oscillations about this steady state, so that the inflaton speed oscillates with a period that we find to be about $4$ e-folds in our simulation, and bursts of gauge field production occur at the maxima of the inflaton speed. In the present work we have pointed out and explored the fact that these  bursts induce a characteristic peaked structure in the SGWB spectrum that gives a distinctive signature accross multiple bands of GW frequencies, from pulsar timing arrays, to astrometry to laser interferometry bands.

Furthermore, we observe that the maximal parity violating coupling between the inflaton-axion and the gauge fields induces a strongly chiral SGWB, where the ratio between left-handed and right-handed GW chiralities is both frequency dependent and oscillating. At very low frequencies (near PTA sensitivity), generated in a stage in which the amount of gauge field amplification is increasing with time, the ratio between the two chiralities is (relatively) close to one, while at higher frequencies (near ET/CE sensitivity) it becomes very different from one, with one polarization (depending on the sign of the inflaton-gauge field interacton) always dominating over the other one. At intermediate frequencies (near Gaia and LISA bands) the GW spectral amplitude and chiral ratio is widely fluctuating over several decades in frequency, a signature that could be used to distinguish this SGWB from others, see~\cite{Bartolo:2016ami,Caprini:2018mtu}.

The results presented in Section~\ref{sec:results} concern a very specific potential that has been chosen to illustrate the range of effects that can be achieved while maintaining consistency with CMB observations. However, our numerical apparatus can be applied to a much broader range of potentials and strengths of the inflation-gauge field coupling, leading to a rich phenomenology.

Finally, our analysis assumes the inflaton field to be homogeneous, while we take into account the full backreaction from the whole spectrum of gauge field fluctuations that are resonantly amplified. The overall agreement with previous lattice simulations provides an argument for the validity of this approximation. However, it is possible that more refined future lattice simulations, with a fully inhomogeneous field content and greater dynamical range than the present ones, may find different results due to the highly non-linear dynamics displayed by this system. Even in the cases in which these inhomogeneities do not alter significantly the background dynamics, we expect that their spectrum will also exhibit a peaked structure, as the one of the GWs studied here, that could result in specific mass spectra of primordial black holes.

\vskip.25cm
\section*{Acknowledgements} 

J.G.-B. acknowledges support from the Research Project PID2021-123012NB-C43 [MICINN-FEDER], and the Centro
de Excelencia Severo Ochoa Program CEX2020-001007-S at IFT.
A.P. is supported by IBS under the project code, IBS-R018-D1. M.P. is supported by Istituto Nazionale di Fisica Nucleare (INFN) through the Theoretical Astroparticle Physics (TAsP) and the Inflation, Dark Matter and the Large-Scale Structure of the Universe (InDark) project. The work of L.S. is partially supported by the US-NSF grant PHY-2112800.

\vskip.25cm

\appendix

\section{Code variables}
\label{app:codevar}

In this appendix we present the rescaled variables and equations that we integrate numerically. As ``time variable'' of our evolutions the number of e-folds is used, defined as 
\begin{equation}
N \equiv \int_{t_0}^t d t' \, H \left( t' \right) \;\;, 
\end{equation}
where $H \equiv \frac{\dot{a}}{a}$ is the Hubble rate, and where $t_0$ is some reference initial time, that we take when the CMB modes are still well inside the horizon. We normalize the scale factor to $1$ at this initial time, so that 
\begin{equation}
a = {\rm e}^N \;\;,\;\; 
\tau = \tau_0 + \int_{t_0}^t \frac{d t'}{a \left( t' \right)} = \tau_0 + \int_0^N \frac{d N}{H \left( N' \right) \, {\rm e}^{N'}} \;. 
\label{def-N}
\end{equation} 

We denote by $V_0$ the value of the inflaton potential at this initial time, and we introduce the dimensionless quantities and fields 
\begin{eqnarray}
&& {\tilde k} \equiv \frac{M_p}{\sqrt{V_0}} \, k \;\;\;,\;\;\; 
{\tilde H} \equiv  \frac{M_p}{\sqrt{V_0}} \, H \;\;\;,\;\;\; 
{\tilde f} \equiv \frac{f}{M_p} \;\;\;,\;\;\; 
{\tilde \phi} \equiv \frac{\phi}{f} \;\;\;,\;\;\; 
{\tilde V} \left( {\tilde \phi} \right) \equiv \frac{V \left( \phi \right)}{V_0} \;, \nonumber\\ 
&& {\bar A}_\pm \equiv \sqrt{2 k} \, {\rm e}^{i k \left( \tau - \tau_0 \right)} \, A_\pm = 
\frac{\sqrt{2 M_p \, {\tilde k}}}{V_0^{1/4}} \, {\rm e}^{i {\tilde k} \int_0^N \frac{d N'}{{\tilde H} \left( N' \right) \, {\rm e}^{N'}}} \, A_\pm \;, 
\label{cod-var}
\end{eqnarray} 
where the rescaling of the gauge field mode eliminates their fast evolving time-dependent phase in the UV regime. In these variables, the last two eqs.~of (\ref{eom-phi-A}) and eq.~(\ref{eq-Ak}) can be combined to give 
\begin{eqnarray}
\frac{d^2 {\tilde \phi}}{d N^2} &=& - \frac{d {\tilde \phi}}{d N} + \frac{{\tilde f}^2}{6} \left( \frac{d {\tilde \phi}}{d N} \right)^3 - \frac{{\tilde V}' \left( {\tilde \phi} \right)}{{\tilde f}^2 {\tilde H}^2} - \frac{2 {\tilde V} \left( {\tilde \phi} \right)}{3 {\tilde H}^2} \, \frac{d {\tilde \phi}}{d N} + \frac{\left\langle \vec{E} \cdot \vec{B} \right\rangle_{\rm symm}}{{\tilde f}^2 V_0 {\tilde H}^2} \;, \nonumber\\ 
\frac{d {\tilde H}}{d N} &=& - 2 {\tilde H} - \frac{{\tilde f}^2 {\tilde H}}{6} \left( \frac{d {\tilde \phi}}{d N} \right)^2 + \frac{2 {\tilde V} \left( {\tilde \phi} \right)}{3 {\tilde H}} \;, \nonumber\\ 
\frac{d^2 {\bar A}_\pm}{d N^2} &=& \left[ 1 + \frac{{\tilde f}^2}{6} \left( \frac{d {\tilde \phi}}{d N} \right)^2 + \frac{2 i {\tilde k}}{{\tilde H} {\rm e}^N} - \frac{2 {\tilde V} \left( {\tilde \phi} \right)}{3 {\tilde H}^2} \right] \frac{d {\bar A}_\pm}{d N} \pm \frac{{\tilde k} {\bar A}_\pm}{{\tilde H} {\rm e}^{N}} \frac{d {\tilde \phi}}{d N} \;,  
\label{code-eqs-phi-A}
\end{eqnarray} 
where prime on the rescaled inflaton potential denotes derivative with respect to the rescaled inflaton. In these variables, the gauge correlators read 
\begin{eqnarray} 
\left\langle \vec{E} \cdot \vec{B} \right\rangle_{\rm symm} &=& -\frac{V_0^2}{8 \pi^2 M_p^4} \, \frac{\tilde H}{{\rm e}^{3 N}} \int d {\tilde k} \, {\tilde k}^2 \sum_\lambda \, \lambda \, \frac{d}{d N} \left\vert {\bar A}_\lambda \right\vert^2 \;, \nonumber\\
\left\langle \frac{\vec{E}^2+\vec{B}^2}{2} \right\rangle_{\rm symm} &=& \frac{V_0^2}{8 \pi^2 M_p^4 \, {\rm e}^{4 N}} \, \int d {\tilde k} \, {\tilde k} \sum_\lambda \left[ {\rm e}^{2 N} {\tilde H}^2 \left\vert  \frac{d {\bar A}_\lambda}{d N} - i \frac{{\tilde k} \, {\bar A}_\lambda}{{\tilde H} \, {\rm e}^N} \right\vert^2 + {\tilde k}^2 \left\vert {\bar A}_\lambda \right\vert^2 \right] \;. 
\label{electro-vevs}
\end{eqnarray} 
In principle, one could subtract the vacuum contribution from the second of (\ref{electro-vevs}), while this is not needed in the first equation, since the vacuum contributions of the two polarizations cancel against each other. We note however that the second expression does not appear in the system (\ref{code-eqs-phi-A}), and it is not used in our numerical integration. 

We integrate the set of equations (\ref{code-eqs-phi-A}) numerically. At each time step, the backreaction integral in the first line of (\ref{electro-vevs}) is evaluated by discretizing the momenta as discussed in Subsection \ref{sec:discretization}, and by using the trapezoid rule. Coherently with the sampling (\ref{sampling}), we discretize using the logarithm of $\tilde{k}_i$ so that the step is identical for all points 
\begin{equation}
    \int_{\tilde{k}_{\rm min}}^{\tilde{k}_{\rm max}} {\cal I}\left(\tilde{k}\right)\;d \tilde{k}\Rightarrow \sum_{i=1}^{i_{\rm max}-1} \; \Delta\ln\tilde{k}_i \frac{\tilde{k}_i{\cal I}_{\tilde{k}_i}+\tilde{k}_{i+1}{\cal I}_{\tilde{k}_i+1}}{2}=\frac{\Delta\ln \tilde{k}}{2}\left(\tilde{k}_{\rm min} {\cal I}_{\tilde{k}_{min}}+\tilde{k}_{\rm max} {\cal I}_{\tilde{k}_{max}}+2 \sum_{i=1}^{i_{\rm max}-1}\tilde{k}_i {\cal I}_{\tilde{k}_{\rm i}}\right) \;.
    \label{eq:integrand-constant-lnk}
\end{equation}
The first two terms in the parenthesis are further disregarded because the integrand assumes negligible values there (otherwise our result would depend on the choice of UV and IR cutoffs). 

We note from the system (\ref{code-eqs-phi-A}) that we chose to integrate a first order differential equation for the Hubble rate, by differentiating the third of (\ref{eom-phi-A}). That equation is then used to set the initial condition for the Hubble rate. More precisely, we choose the initial time well before the CMB modes left the horizon, and with a sufficiently flat inflaton potential, such that the inflaton is performing a standard slow-roll evolution, with negligible gauge field amplification and backreaction. The gauge modes are initially in the adiabatic vacuum, and the full set of initial conditions at $N =0$ is given by 
\begin{equation}
{\tilde \phi} \left( 0 \right) = {\tilde \phi}_0 \;\;,\;\; {\tilde H} \left( 0 \right) = \frac{1}{\sqrt{3}} \;\;,\;\; \frac{d {\tilde \phi}}{d N} \Big\vert_{N=0} = - \frac{1}{{\tilde f}^2}  \, \frac{d {\tilde V}}{d {\tilde \phi}} \Big\vert_{{\tilde \phi} = {\tilde \phi}_0} \;\;,\;\; {\bar A}_\lambda \left( 0 \right) = 1 \;\;,\;\; \frac{d {\bar A}_\lambda}{d N}  \Big\vert_{N=0} = 0 \;.
\label{initial1}
\end{equation} 

We now discuss the set of equations that we integrate numerically to evaluate the GW production. Rewritten in terms of $N$, eq.~(\ref{eom_Qlambda}) rewrites
\begin{equation}
{\hat O}_N \, {\hat Q}_\lambda \left( N ,\, \vec{k} \right) = \frac{\hat{\cal S}_\lambda \left( N,\,\vec{k} \right)}{H^2 \, {\rm e}^{2 N}} \;, 
\end{equation}
where we introduced the operator 
\begin{equation}
{\hat O}_N \equiv \frac{d^2}{d N^2} + \left[ 1 + \frac{1}{H} \frac{d H}{d N} \right] \frac{d}{d N} + \left[ \frac{k^2}{H^2 \, {\rm e}^{2 N} } - 2   - \frac{1}{H} \frac{d H}{d N}  \right] \;. 
\end{equation}

In terms of the Green function 
\begin{equation}
{\hat O}_N {\cal G}_N \left( N ,\, N' \right) = \delta \left( N - N' \right) \;\;\;,\;\;\; 
{\cal G}_N \left( N ,\, N' \right) = \tilde{\cal G}_N \left( N ,\, N' \right) \, \Theta \left( N - N' \right) \;, 
\label{green-N}
\end{equation} 
we have the formal solution 
\begin{equation} 
\hat{Q}_\lambda \left( \vec{k},\, N \right) = \int_0^N d N' \tilde{\cal G}_k \left( N ,\, N' \right) \, \frac{ \hat{\cal S}_\lambda(N,\,\vec{k}) }{\left( H \left( N' \right) \, {\rm e}^{N'} \right)^2} \;. 
\end{equation} 
As this solution coincides with (\ref{Q1lambda_formal}), rewritten in terms of $N$ instead of $\tau$, we can immediately relate the Green functions in the two variables 
\begin{equation}
\tilde{\cal G}_k \left( N ,\, N' \right) = H \left( N' \right) \, {\rm e}^{N'} \, {\tilde G}_k \left( \tau \left( N \right) ,\, \tau' \left( N' \right) \right) \;, 
\end{equation}
where the relation between $\tau$ and $N$ is given in (\ref{def-N}). The explicit expression for the Green function in code variables is given at the end of this appendix. 

We then proceed as in Subsection \ref{subsec:sourced-GW} of the main text. In particular, eq.~(\ref{P-la-main}) rewrites~\footnote{As $\tilde{\cal G}_k$ is dimensionless, the second line of this expression has mass dimension $-2$, so that the full result is dimensionless, as it should be.}
\begin{eqnarray} 
&& 
P_\lambda \left( k \right)  =  \frac{k^3  {\rm e}^{-2N} }{16 \pi^4 M_p^4} 
\; \int_0^\infty d p \, p^2 \int_{-1}^1 d \cos \theta 
\left( 1 - \lambda \frac{p \cos \theta - k}{\sqrt{k^2 -2 k \, p_* \cos \theta + p_*^2}} \right)^2 
\left( 1 + \lambda \cos \theta \right)^2 \nonumber\\ 
&& 
\times\left\vert \int_0^N d N' \frac{\tilde{\cal G}_k \left( N ,\, N' \right) }{\rm e^{N'}} 
\left[ 
\frac{d A_+ \left( N' ,\, p \right)}{d N'}  \frac{d A_+ \left( N' ,\, \vert \vec{k} - \vec{p} \vert \right)}{d N'}  + 
\frac{p \, \vert \vec{k} - \vec{p} \vert}{H^2 \left( N' \right) {\rm e}^{2 N'} } \,  A_+ (N',\, p)  \,  A_+ (N',\, \vert \vec{k} - \vec{p} \vert) 
\right] \right\vert^2 \;, \nonumber\\ 
\end{eqnarray} 
where we have performed one trivial angular integration (the rotation of the integration variable $\vec{p}$ at fixed angle $\theta$ with $\vec{k}$). To perform the remaining two integrals we introduce the variables $p \equiv k \left( X + Y \right) \,,\; 
q \equiv \left\vert \vec{k} - \vec{p} \right\vert \equiv k \left( X - Y \right)$. In terms of these variable, the integral of any function $f$ of $p$ and of $\cos \theta$ reads 
\begin{equation}
\int_0^\infty d p\, p^2 \int_{-1}^1 d \cos \theta \; f \left( p ,\, \cos \theta \right) = 2 k^3 \int_{1/2}^\infty d X \int_{-1/2}^{+1/2} d Y \, \left( X^2 - Y^2 \right) \; f \left( k \left( X + Y \right) ,\, \frac{1+4 X Y}{2 \left( X + Y \right)} \right) \;. 
\label{pth2XY}
\end{equation} 
The momentum integral leading to $P_\lambda \left( k \right)$ can be visualized as a one loop integral with internal momenta $\vec{p}$ and $\vec{q} = \vec{k} - \vec{p}$. In the boundary of (\ref{pth2XY}) the two momenta $\vec{p}$ and $\vec{q}$ are either aligned or anti-aligned. More specifically, let us consider the $\left\{ X ,\, Y \right\} = \left\{ \frac{1}{2} ,\, - \frac{1}{2} \right\}$  ``corner'', corresponding to $\vec{p}=0$ and $\vec{q} = \vec{k}$. Starting from this point and moving along the $Y = - \frac{1}{2}$ boundary corresponds to taking $\vec{p}$ and $\vec{q}$ anti-aligned, with $\vec{p}$ pointing in the opposite direction to $\vec{k}$ (namely, $\cos \theta = - 1$), and $q > p$. The two remaining boundaries have instead $\vec{p}$ pointing and in the same direction as $\vec{k}$ (namely, $\cos \theta = + 1$). In the $X = \frac{1}{2}$ segment, both $\vec{p}$ and $\vec{q}$ are directed as $\vec{k}$, while in the remaining $Y = \frac{1}{2}$ boundary $\vec{q}$ is directed opposite to the other two vectors, with $p > q$. These two boundaries join each other at the $\left\{ X ,\, Y \right\} = \left\{ \frac{1}{2} ,\, \frac{1}{2} \right\}$  ``corner'', where $\vec{p} = \vec{k}$, while $\vec{q}$ vanishes. 

In these variables we obtain 
\begin{eqnarray} 
&& \!\!\!\!\!\!\!\!  \!\!\!\!\!\!\!\!  \!\!\!\!\!\!\!\!  \!\!\!\!\!\!\!\!  
P_\lambda \left( k \right)  =  \frac{k^6  {\rm e}^{-2N} }{128 \pi^4 M_p^4} 
\;  \int_{1/2}^{\infty} d X \int_{-1/2}^{1/2} d Y \, 
\frac{\left( 1 - 4 Y^2 \right)^2 \left( 1 + 2 \lambda X \right)^4}{X^2-Y^2} \; \Bigg\vert \int_0^N d N' \frac{\tilde{\cal G}_k \left( N ,\, N' \right)}{{\rm e}^{N'}} 
 \nonumber\\ 
&&  \!\!\!\!\!\!\!\!  \!\!\!\!\!\!\!\! 
\left[ \frac{d A_+ \left( N' ,\, k \left( X + Y \right) \right)}{d N'}  \frac{d A_+ \left( N' ,\, k \left( X - Y \right) \right)}{d N'}  + \frac{k^2 \, \left( X^2-Y^2 \right)}{H^2 \left( N' \right) {\rm e}^{2 N'} } \,  A_+ (N',\, k \left( X + Y \right))  \,  A_+ (N',\, k \left( X - Y \right)) 
\right] \Bigg\vert^2 \;, \nonumber\\ 
\end{eqnarray} 
and we note that the integrand is symmetric under $Y \to - Y$, which corresponds to interchanging the internal momenta $\vec{p} \leftrightarrow \vec{q}$ in the loop. Performing the rescalings (\ref{cod-var}) finally results in 
\begin{eqnarray} 
P_\lambda \left( k \right) & = & \frac{{\rm e}^{-2N} V_0^2}{512 \pi^4 M_p^8} \;  \int_{1/2}^{\infty} d X \int_{-1/2}^{1/2} d Y \, 
\frac{\left( 1 - 4 Y^2 \right)^2 \left( 1 + 2 \lambda X \right)^4}{\left( X^2-Y^2 \right)^2} 
\; \Bigg\vert \int_0^N d N' \; \frac{{\tilde k}^2 \, \tilde{\cal G}_k \left( N ,\, N' \right) }{\rm e^{N'}} 
 \nonumber\\ 
&& \times \, 
\Bigg\{ \frac{d {\bar A}_+ \left( N' ,\, {\tilde k} \left( X + Y \right) \right)}{d N'} 
\frac{d {\bar A}_+ \left( N' ,\, {\tilde k} \left( X - Y \right) \right)}{d N'}  \nonumber\\ 
&& \quad\quad \quad\quad - \frac{i \; {\tilde k}}{{\rm e}^{N'} \, {\tilde H} \left( N' \right)} \Bigg[ 
\left( X + Y \right) {\bar A}_+ \left( N' ,\, {\tilde k} \left( X + Y \right) \right) 
\frac{d {\bar A}_+ \left( N' ,\, {\tilde k} \left( X - Y \right) \right)}{d N'}
\nonumber\\ 
&& \quad\quad\quad\quad   \quad\quad\quad\quad \quad\quad 
+  \left( X - Y \right) \frac{d {\bar A}_+ \left( N' ,\, {\tilde k} \left( X + Y \right) \right)}{d N'} 
{\bar A}_+ \left( N' ,\, {\tilde k} \left( X - Y \right) \right) \Bigg] \Bigg\} \Bigg\vert^2 \nonumber\\ 
&\equiv& \int_{1/2}^{\infty} d X \int_{-1/2}^{1/2} d Y \, {\cal C} \left( X ,\, Y \right) \;. 
\label{code-N-PS}
\end{eqnarray} 

We conclude this section by presenting the Green function in code variables. The Green function can be obtained following the steps outlined in Subsection \ref{subsec:Green} of the main text, by replacing the differential operator in eq.~(\ref{conditions-Green}) by that in eq.~(\ref{green-N}). The first step is to solve the homogeneous differential equation ${\hat O}_N {\cal F} \left( N ,\, k \right) = 0$ subject to the sub-horizon initial condition (\ref{adibatic-in}), rewritten in terms of the variable $N$. To rescale away this fast evolving initial phase, we rescale the wanted solution as done in eq.~(\ref{cod-var}) for the gauge modes 
\begin{equation}
\bar{\cal F} \left( N ,\, k \right) \equiv \sqrt{2 k} \, {\rm e}^{i k \left( \tau - \tau_0 \right)} \,{\cal F} \left( N ,\, k \right) = 
\frac{\sqrt{2 M_p \, {\tilde k}}}{V_0^{1/4}} \, {\rm e}^{i {\tilde k} \int_0^N \frac{d N'}{{\tilde H} \left( N' \right) \, {\rm e}^{N'}}} \, {\cal F} \left( N ,\, k \right)  \;, 
\end{equation}
where the rescaled function is fully determined by~\footnote{As a check of our the numerical scheme, we actually implemented the initial conditions for ${\bar A}_\pm$ and $\bar{\cal F}$ including the first order term in ${\tilde H} \left( 0 \right) / {\tilde k}$, and we verified that, for the initial times that we chose, these corrections produced negligible effects on the final results.}
\begin{eqnarray}
\left\{ \begin{array}{l} 
\frac{d^2 \, \bar{\cal F} \left( N ,\, k \right)}{d N^2} 
+ \left[ 1 + \frac{1}{\tilde H} \frac{d {\tilde H}}{d N} - \frac{2 i \, {\tilde k}}{{\tilde H} \, {\rm e}^N} \right] \frac{d \, \bar{\cal F} \left( N ,\, k \right)}{d N} 
- \left[  2 + \frac{1}{\tilde H} \frac{d {\tilde H}}{d N}  \right] \, \bar{\cal F} \left( N ,\, k \right) = 0 \;, \\ \\ 
\bar{\cal F} \big\vert_{N=0} = 1 \;\;\;,\;\;\; 
\frac{d \bar{\cal F}}{d N} \big\vert_{N=0} = 0 \;. 
\end{array} \right. 
\label{eqs-N-barF}
\end{eqnarray} 
Once $\bar{\cal F} \left( N ,\, k \right)$ is determined numerically, we employ it in the Green function 
\begin{eqnarray} 
\tilde{\cal G}_k \left( N ,\, N' \right) &=& 
\frac{{\rm Im } \left[ {\cal F} \left( N ,\, k \right) {\cal F}^* \left( N' ,\, k \right) \right]}{{\rm Im } \left[ \frac{d {\cal F} \left( N' ,\, k \right)}{d N'} {\cal F}^*\left( N' ,\, k \right) \right]} \nonumber\\ 
& = & \frac{{\rm Im } \left[ {\rm e}^{- i {\tilde k}  \int_{N'}^N \frac{d n}{{\tilde H} \left( n \right)\, {\rm e}^n }  } \, \bar{\cal F}  \left( N ,\, k  \right) \, \bar{\cal F}^*  \left( N' ,\, k  \right) \right]}{{\rm Im } \left[ \frac{d \, \bar{\cal F} \left( N' ,\, k \right)}{d N'} \,  \bar{\cal F}^*  \left( N' ,\, k  \right) \right] - \frac{\tilde k}{{\tilde H} \left( N' \right) \, {\rm e}^{N'}} \left\vert \bar{\cal F} \left( N' ,\, k  \right) \right\vert^2 } \;. 
\label{Green-N-formalsol}
\end{eqnarray} 

As derived analytically in eq.~(\ref{Green-tau-super}), and as we verify numerically, in the super-horizon regime the Green function satisfies 
\begin{eqnarray}
\lim_{{\tilde k} \ll {\rm e}^N  H \left( N \right), \, {\rm e}^{N'}  H \left( N' \right)} \tilde{\cal G}_k \left( N ,\, N' \right) &=& {\tilde H} \left( N' \right) \, {\rm e}^{N+2N'} \int_{N'}^N \frac{d \tau''}{{\tilde H} \left( N'' \right) {\rm e}^{3 N''}} \;. 
\label{Green-N-super}
\end{eqnarray}

\section{Results at constant $\xi$}
\label{app:constxi}

In this Appendix we evaluate some of the expressions of the main text in the case of constant $\xi$, to compare with some results obtained in the literature in this case. 

We start from the gauge field mode functions and correlators. For constant $\xi$, and for a de Sitter background, namely, to zeroth order in slow-roll, eq.~(\ref{def-xi}) is solved exactly by an irregular Coulomb function. Ref.~\cite{Anber:2009ua} provided a simpler expression, 
\begin{equation}
A_+ \simeq \frac{1}{\sqrt{2 k}} \left( \frac{-k \tau}{2 \xi} \right)^{1/4} \, {\rm e}^{\pi \xi - 2 \sqrt{-2 \xi k \tau}} \;, 
\label{A-xiconst}
\end{equation} 
that approximates very well the exact solution at its maximum, which is what matters in computing the effects of the gauge field amplification (see ref.~\cite{Peloso:2016gqs} for a detailed discussion). Correspondingly, the $E-$ and $B-$mode functions defined in eq.~(\ref{F-EB}) are 
\begin{equation}
F_E \simeq - \sqrt{\frac{2 \xi}{-k \tau}} \, A_+ 
\;\;\;,\;\;\; 
F_B = k \, A_+ \;, 
\label{FEB-xiconst}
\end{equation}
and the correlators (\ref{CE-CB}) evaluate to 
\begin{eqnarray} 
&& \left\langle E_i \, E_i \right\rangle_{\rm symm} \simeq 2.78 \times 10^{-4} H^4 \, \frac{{\rm e}^{2 \pi \xi}}{\xi^3} \;\;,\;\; 
\left\langle B_i \, B_i \right\rangle_S = 2.44 \times 10^{-4} H^4 \, \frac{{\rm e}^{2 \pi \xi}}{\xi^5} \ll \left\langle {\hat E}_i \, {\hat E}_i \right\rangle_{\rm symm} \;, \nonumber\\ 
&& \left\langle E_i \, B_i \right\rangle_{\rm symm} = -2.44 \times 10^{-4} H^4 \, \frac{{\rm e}^{2 \pi \xi}}{\xi^4} \;, 
\end{eqnarray} 
in agreement with eq.~(2.15) of \cite{Barnaby:2011vw}. The inequality in the first line holds in the phenomenologically relevant regime $\xi \gg 1$. 

Next, we discuss the GWs production in a de Sitter geometry and with constant $\xi$. We start from eq.~(\ref{P-la-main}) of the main text, where we recall that, for the dS case, the Green function evaluated for super-horizon modes reads (compare with eq~(\ref{Green-dS}))
\begin{equation}
\lim_{\tau \to 0} {\tilde G}_k \left( \tau ,\, \tau' \right) = \frac{k \tau' \cos \left( k \tau' \right) - \sin \left( k \tau' \right)}{k^3 \tau \tau'} \;. 
\label{green-dS-tau-super}
\end{equation} 

We insert this expression in eq.~(\ref{P-la-main}), together with the mode functions (\ref{A-xiconst}) and (\ref{FEB-xiconst}), and, after some lengthy but straightforward algebra, we arrive to 
\begin{eqnarray} 
&& \!\!\!\!\!\!\!\!  \!\!\!\!\!\!\!\!  \!\!\!\!\!\!\!\! 
P_\lambda \left( k \right)  = \frac{H^4}{2 \pi^2 M_p^4} \frac{{\rm e}^{4 \pi \xi}}{\xi}   \nonumber\\ 
&& \!\!\!\!\!\!\!\!  \!\!\!\!\!\!\!\! 
\times 
\int \frac{d^3 p_* }{\left( 2 \pi \right)^3} 
p_*^{1/2} \, \vert {\hat k} - \vec{p}_* \vert^{1/2} 
\frac{\left( 1 + \lambda \cos \theta \right)^2  \left(  1 - p_* \cos \theta  + \lambda \sqrt{1-2 p_* \cos \theta + p_*^2} \right)^2}{16 \left( 1-2 p_* \cos \theta + p_*^2 \right)} \nonumber\\ 
&& 
\left\{ \int_{-k \tau}^\infty d x x^{1/2} \left[ \sin x - x \cos x \right] 
\left[ \frac{2 \xi}{x} + \sqrt{p_* \, \vert {\hat k} - \vec{p}_* \vert} \right] {\rm e}^{- 2 \sqrt{2 \xi x } \left[ \sqrt{p_*} + \sqrt{\vert {\hat k} - \vec{p}_* \vert} \right]}  \right\}^2 \;, 
\label{Pla-app-constxi1}
\end{eqnarray} 
where $\vec{p}_* \equiv \vec{p} / k$. The second line contains the two integrals 
\begin{eqnarray}
I_1 &\equiv& 2 \xi \int_0^\infty d x x^{-1/2} \left( \sin x - x \cos x \right) {\rm e}^{-2 Q \sqrt{2 \xi x}} \;,
\nonumber\\ 
I_2 &=& \sqrt{p_* \, \vert {\hat k} - \vec{p}_* \vert}  \int_0^\infty d x x^{1/2} \left( \sin x - x \cos x \right) {\rm e}^{-2 Q \sqrt{2 \xi x}} \;, 
\end{eqnarray} 
where $Q \equiv \left[ \sqrt{p_*} + \sqrt{\vert {\hat k} - \vec{p}_* \vert} \right]$, and where we set $- k \tau \to 0$ in the lower extremum of integration, which is appropriate for the solution in the super-horizon regime. At $\xi \gg 1$ we can approximate 
\begin{eqnarray} 
I_1 &\simeq& \frac{2 \xi}{3} \int_0^\infty d x \, x^{5/2} \,  {\rm e}^{-2 Q \sqrt{2 \xi x}} 
= \frac{2 \xi}{3} \, \frac{45}{32 \sqrt{2} Q^7 \xi^{7/2}} \;, \nonumber\\ 
I_2 &\simeq& \frac{1}{3} \sqrt{p_* \, \vert {\hat k} - \vec{p}_* \vert}  \int_0^\infty d x\, x^{7/2} {\rm e}^{-2 Q \sqrt{2 \xi x}} = \frac{1}{3} \sqrt{p_* \, \vert {\hat k} - \vec{p}_* \vert} \frac{315}{32 \sqrt{2} Q^9 \xi^{9/2}} \;. 
\end{eqnarray} 

We insert these expressions into eq.~(\ref{Pla-app-constxi1}) and perform the $d^3 p_*$ integration numerically (in fact one angular integration is trivial, as we also discussed in the previous appendix). We verified that $I_2$ provides a negligible contribution in the $\xi \gg 1$ regime (being  suppressed by one additional power of $1/\xi$ relative to $I_1$). The final result for the left handed ($\lambda = +1$) and right handed  ($\lambda = -1$) GWs is 
\begin{equation}
P_L \left( k \right) \simeq \frac{8.72 \cdot 10^{-8} H^4 \,{\rm e}^{4 \pi \xi}}{M_p^4\,\xi^6} \;\;,\;\; 
P_R \left( k \right) \simeq \frac{1.86 \cdot 10^{-10} H^4 \,{\rm e}^{4 \pi \xi}}{M_p^4\, \xi^6} \;,  
\label{eq:power-constant-xi}
\end{equation} 
which agrees with eqs.~(3.39) and~(3.41) of ref.~\cite{Barnaby:2011vw}. We note that our eq.~(\ref{Pla-app-constxi1}) is a factor of $4$ smaller than eq.~(3.40) of ref. \cite{Barnaby:2011vw}. Given the agreement between their and our final numerical values, we conclude that eq.~(3.40) of ref.~\cite{Barnaby:2011vw} is affected by a typo.

\section{Numerical implementation of the UV cutoffs}
\label{app:numerics1}

We devote this appendix to a detailed explanation of our implementation of the momentum cutoff of the backreaction integral and the equations of motion of the gauge field and auxiliary function $\tilde{\cal F}$. 

Regarding the backreaction cutoff, we define the regularized integral as follows
\begin{eqnarray}
    {\cal I}_{\rm reg}(\tilde{k})\equiv R(N,\tilde{k})\,{\cal I}(\tilde{k})\;,
\end{eqnarray}
where $R(N,\tilde{k})$ is a regularization function. For all practical purposes this function could take the form of a Heaviside step function, but in practice the numerical differential equation solver in \textit{Mathematica} works better with analytic functions. As a result we choose the following function for the regulator
\begin{equation}
    R(N,\tilde{k})\equiv\frac{1}{2}\left\{\tanh\left[10^3\ln\left(\frac{\tilde{k}_{\rm reg}(N)}{\tilde{k}}\right)\right]+1\right\} \;,
    \label{eq:regulator}
\end{equation}
where the number $10^3$ controls the steepness of the regulator and in practice our choice makes it indistinguishable from a Heaviside step function. The function $\tilde{k}_{\rm reg}$ defined in (\ref{eq:increasing-cutoff}) has to be solved for numerically along with the equations of motion. In the initial stages of inflation while the system is still in the low backreaction regime, $\tilde{k}_{\rm thr}(N)$ is a monotonously increasing function of time. During this period, the momentum cutoff and the threshold between stability and instability are identical, 
\begin{eqnarray}
    \tilde{k}_{\rm reg}(N)=\tilde{k}_{\rm thr}(N), \;\;\;\; {\rm early \;times \;and\; negligible \; backreaction} \;.
\end{eqnarray}
Subsequently, when the system enters the strong backreaction regime, the function $\tilde{k}_{\rm thr}(N)$ evolves in a non-monotonous way; however, the momentum cutoff $\tilde{k}_{\rm reg}(N)$ should be a monotonously increasing function of the number of e-folds that is equal to the greatest value that $\tilde{k}_{\rm thr}(N)$ has ever obtained before this moment. We implement this using \textit{event triggers} in \textit{Mathematica}. 

During the evolution of the numerical differential solver \textbf{NDSolve}, we define an event trigger using the function \textbf{WhenEvent} which checks at every instant in time whether an event satisfying a condition has been triggered. The condition in our case is the change in the monotonicity of $\tilde{k}_{\rm thr}(N)$, which can also be rewritten as 

\begin{eqnarray}
    \frac{d \tilde{k}_{\rm thr}}{dN}= {\rm e}^{N}\tilde{H}\frac{d\tilde{\phi}}{dN}+ {\rm e}^{N}\frac{d\tilde{H}}{dN}\frac{d\tilde{\phi}}{dN}+ {\rm e}^{N}\tilde{H}\frac{d^2\tilde{\phi}}{dN^2} < 0 \;.
\end{eqnarray}
This implies that the event will trigger once the derivative of $\tilde{k}_{\rm thr}$ switches from positive to negative. When this event triggers, we program \textit{Mathematica} to save the value of $\tilde{k}_{\rm thr}$ at this particular moment and use the function \textbf{AppendTo} to append the value of $\tilde{k}_{\rm thr}$ in an initially empty list which we call $\tilde{\cal{K}}_{\rm loc,max}$, therefore saving the local maxima of $\tilde{k}_{\rm thr}$ as the evolution progresses. We use the option \textbf{DiscreteVariables} to include in the evolution a discrete variable which is a function to be evaluated by \textbf{NDSolve}, that has an initial condition but that does not obey a corresponding differential equation. We call this function $\tilde{k}_{\rm loc,max}(N)$, and the code changes its value at every event trigger to be $\tilde{k}_{\rm loc,max}\rightarrow {\rm MAX}\left[\tilde{\cal{K}}_{\rm loc,max}\right]$. 

\begin{figure}
    \centering
    \includegraphics[width=0.66\linewidth]{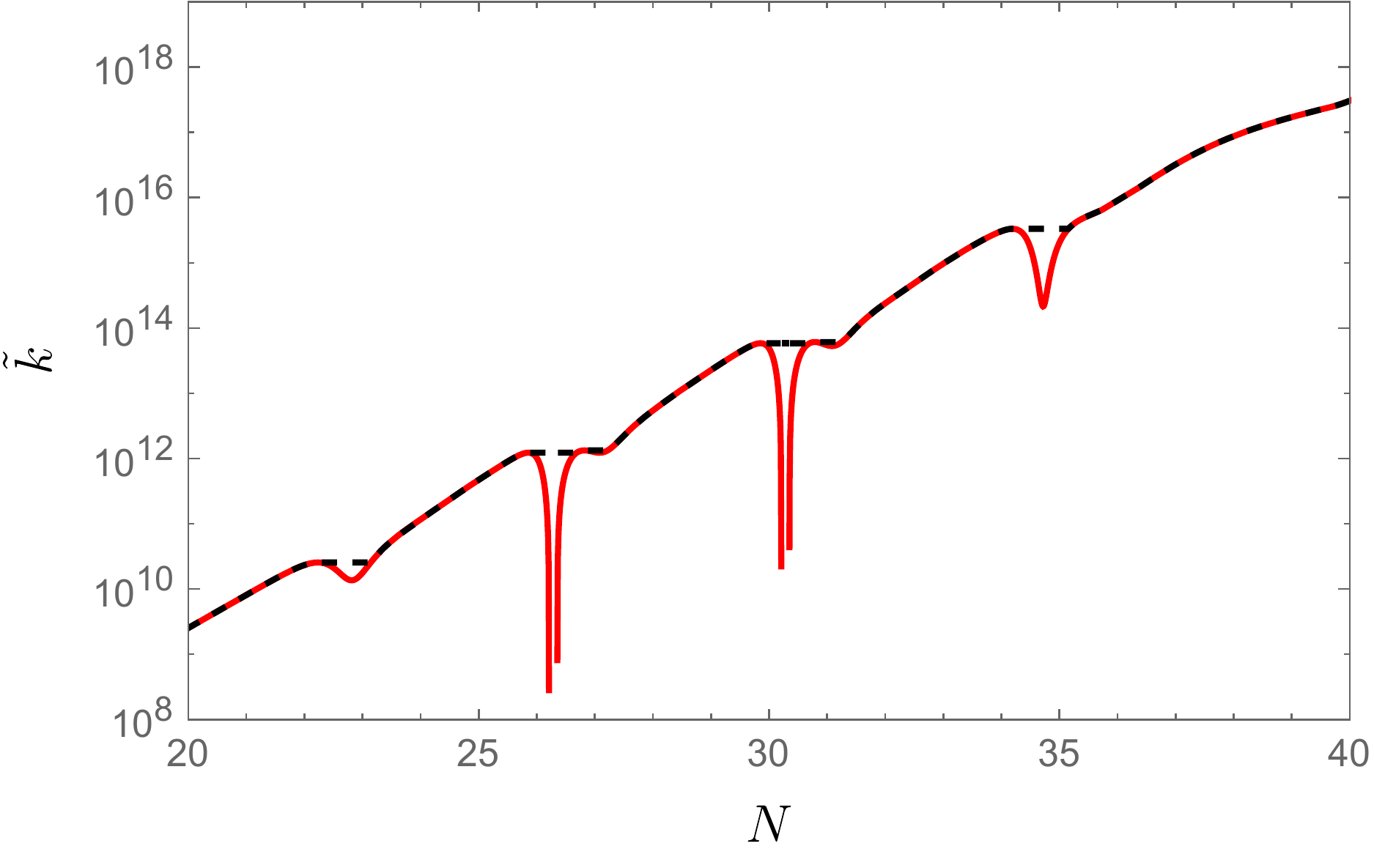}
    \caption{Comparison between $\tilde{k}_{\rm thr}(N)$ and $\tilde{k}_{\rm reg}(N)$ for the scenario presented in section \ref{sec:results}. The black dashed line is given by (\ref{eq:kreg}) whereas the red line is the threshold between stability and instability defined in (\ref{eq:kthr}) The red lines vanish briefly because the direction of motion of the inflaton switches momentarily and the plot does not have enough resolution to keep track of the rapidly decreasing value in logarithmic units. Clearly, this has no effect on the two regulators ${\hat k}_{\rm reg} \left( N \right)$ and ${\hat k}_{\rm vac} \left( N \right)$, that are sensitive only to the maximum value attained by $\tilde{k}_{\rm thr}$ up to $N$. 
    }
    \label{fig:kreg}
\end{figure}

We then define the momentum cutoff as 

\begin{eqnarray}
    \tilde{k}_{\rm reg}(N) &=& \frac{1}{2}\left\{\tanh\left[10^3\ln\left(\frac{\tilde{k}_{\rm thr}(N)}{\tilde{k}_{\rm loc,max}(N)}\right)\right]+1\right\}\tilde{k}_{\rm thr}(N)\nonumber\\
     &&\quad\quad\quad\quad \quad\quad\quad +\frac{1}{2}\left\{\tanh\left[-10^3\ln\left(\frac{\tilde{k}_{\rm thr}(N)}{\tilde{k}_{\rm loc,max}(N)}\right)\right]+1\right\}\tilde{k}_{\rm loc,max}(N) \;. 
     \label{eq:kreg}
\end{eqnarray}
This function effectively picks out the maximum value ever attained by $\tilde{k}_{\rm thr}$, from the start of the code up to the moment $N$. This function is plotted for the parameters chosen in the main body in figure \ref{fig:kreg}. The algorithm described above allows for the computation of the regularized backreaction integral with the correct cutoff in real time without the need for iterative work. 

Let us now turn our attention to the UV cutoff of the evolution of the gauge modes $\bar{A}$ and auxiliary functions $\bar{\cal F}$. The cutoff can be easily implemented by setting the entire right hand side of the last line of (\ref{code-eqs-phi-A}) equal to zero for any comoving momentum above $\tilde{k}_{\rm vac}(N)$ as follows. 
\begin{eqnarray}
    \frac{d^2 {\bar A}_\pm}{d N^2} &=& R(N,\tilde{k}) \left\{\left[ 1 + \frac{{\tilde f}^2}{6} \left( \frac{d {\tilde \phi}}{d N} \right)^2 + \frac{2 i {\tilde k}}{{\tilde H} {\rm e}^N} - \frac{2 {\tilde V} \left( {\tilde \phi} \right)}{3 {\tilde H}^2} \right] \frac{d {\bar A}_\pm}{d N} \pm \frac{{\tilde k} {\bar A}_\pm}{{\tilde H} {\rm e}^{N}} \frac{d {\tilde \phi}}{d N}\right\} \;,  
    \label{eq:reg-A}
\end{eqnarray}
with the replacement $\tilde{k}_{\rm reg}(N) \rightarrow \tilde{k}_{\rm vac}(N)$ in the numerator of (\ref{eq:regulator}) and an analogous expression for (\ref{code-N-PS}). This is essentially a “trick” that takes advantage of the fact that in the vacuum configuration the first derivative of the mode functions is formally zero. Since the speed is zero, one only needs to set the second derivative to zero to guarantee that the modes are frozen. Once the threshold $\tilde{k}_{\rm vac}(N)$ grows beyond a specific $\tilde{k}$, the regulator becomes effectively equal to one and the mode starts evolving from the vacuum configuration while obeying the appropriate equation of motion inside the curly bracket of (\ref{eq:reg-A}).

\section{Additional comments on the numerics with \textit{Mathematica}}
\label{app:numerics2}

This appendix is dedicated to providing the reader with a bird's-eye view of the numerical setup with \textit{Mathematica}. Our set of differential equations consists of the scalar field equation of motion, the equation for the Hubble rate, one equation for each gauge mode and auxiliary function $\bar{{\cal F}}$. There are $1+2 i_{\rm max}$ second order and $1$ first order (namely the one for the rescaled Hubble rate) differential equations (note that we ignore in the code the polarization of the gauge field that is not expected to become tachyonic). We solve all equations isolating the greatest derivative order on the left hand side and moving every other term to the right hand side. Subsequently we reduce the order of the system to be first order by introducing auxiliary variables whenever necessary. Specifically, for a generic variable $Z(N)$ which obeys a second order differential equation, we introduce the auxiliary variable $Z_d(N)$ and rewrite the system as follows
\begin{eqnarray}
    Z''(N)=\dots\;\;\;\;\; \rightarrow \;\;\;\;\; \left\{Z_d'(N)=\dots , \;Z'(N)=Z_d(N)\right\}\,.
\end{eqnarray}

This reduces the system to a total of $3+4 i_{\rm max}$ first order differential equations. Subsequently, we separate the real and imaginary parts of the gauge modes and auxiliary functions $\bar{{\cal F}}$ and evolve them separately using the expansion
\begin{eqnarray}
    \bar{A}(N,\tilde{k})=\bar{A}_{\rm RE}(N,\tilde{k})+i \bar{A}_{\rm IM}(N,\tilde{k})\;.
\end{eqnarray}

This step manifestly eliminates every imaginary unit in the equations of motion and generally provides stability to the code since everything is assumed a-priori to be a real variable. The total number of equations is then $3+8 i_{\rm max}$ (the number of equations does not double, since the inflaton, its derivative, and the Hubble rate are real quantities). Finally, adding the auxiliary discrete variable introduced in Appendix~\ref{app:numerics1}, necessary for the computation of the backreaction cutoff, we end up with a total of $3204$ equations for $i_{\rm max}=400$.

We then evolve the system using the function \textbf{NDSolve} and using the method \textbf{Method$\rightarrow$ $\{$"StiffnessSwitching", Method$\rightarrow$ $\{$"ExplicitRungeKutta, Automatic$\}\}$}. This method is ideal for the type of differential equations we are studying since it switches between a stiff and non-stiff solver depending on whether the criteria for stiffness are satisfied. During the low backreaction regimes the code generally evolves using the Runge-Kutta approach which is very fast and efficient and conveniently allows \textit{Mathematica} to make use of autoparallelization, which is a process by which the large summations in our code are vectorized and performed in parallel among all available cores. On the other hand, when the bursts of particle productions occur, \textit{Mathematica} switches to a stiff solver that is slower, but more appropriate to the strong backreaction regime. 

We have tested the robustness of the code by performing trials with varying discretization options and for limited range of modes as well as various accuracy and precision requirements. We also tested various other numerical differential equation solvers and achieved identical results but for a large price in speed and memory efficiency.  

The raw results of the numerical solver are then used in (\ref{Green-N-formalsol}) and (\ref{code-N-PS}) for the computation of the gravitational wave power spectrum. This process involves certain technical aspects that are worth mentioning. We discretize the time variable into $\Delta N=0.5$ intervals and define the real and imaginary part of the gauge modes as a smooth function of momentum by interpolating the solutions over $k_i$ over the fixed time. These smooth functions have a hard UV cutoff which is fixed to be $\tilde{k}_{\rm reg}$. This step regularizes the gravitational wave integral, eliminating any possible unphysical UV contribution.

We then define a grid in $(\ln(X),Y)$ space that consist of $100\times 100$ points equally spaced in the two variables. We choose variables $\ln(X)$ and $Y$ because we observe that the dominant contribution in the $X$ direction may be consolidated anywhere between small $X\sim 1/2$ values or very large $X\sim {\cal O}(100)$ and we want to capture the integrand (\ref{code-N-PS}) accurately at both scales. On the other hand the variable $Y$ is sampled linearly because that is sufficient to capture the dominant contribution to the integrand for all the possible integrand shapes that may arise. 

We then proceed to compute the gravitational wave power for an external momentum $\tilde{k}$ by selecting the mode of interest from our original set of $i_{\rm max}=400$ modes. The one dimensional time integral contained in the absolute value of (\ref{code-N-PS}) is computed first using the one dimensional trapezoid rule, once for each combination of $(\ln(X),Y)$ for a total of $100\times 100$ times. We finally multiply the absolute value square of the integral with the polarization dependent prefactor and compute the integral over $dX$ and $dY$ using the 2-dimensional version of the trapezoid rule separately for each polarization. We also apply the same simplifying identity to the 2-dimensional trapezoid rule as the 1-dimensional version written in (\ref{eq:integrand-constant-lnk}) which in this case dramatically speeds up the integration. We have experimented with the time and momentum spacing and determined that the process described above leads to convergent results.

\section{Momentum dependence of the GW source and implications for the GW polarization}
\label{app:parity}

\begin{figure}
    \centering
    \includegraphics[width=0.45\linewidth]{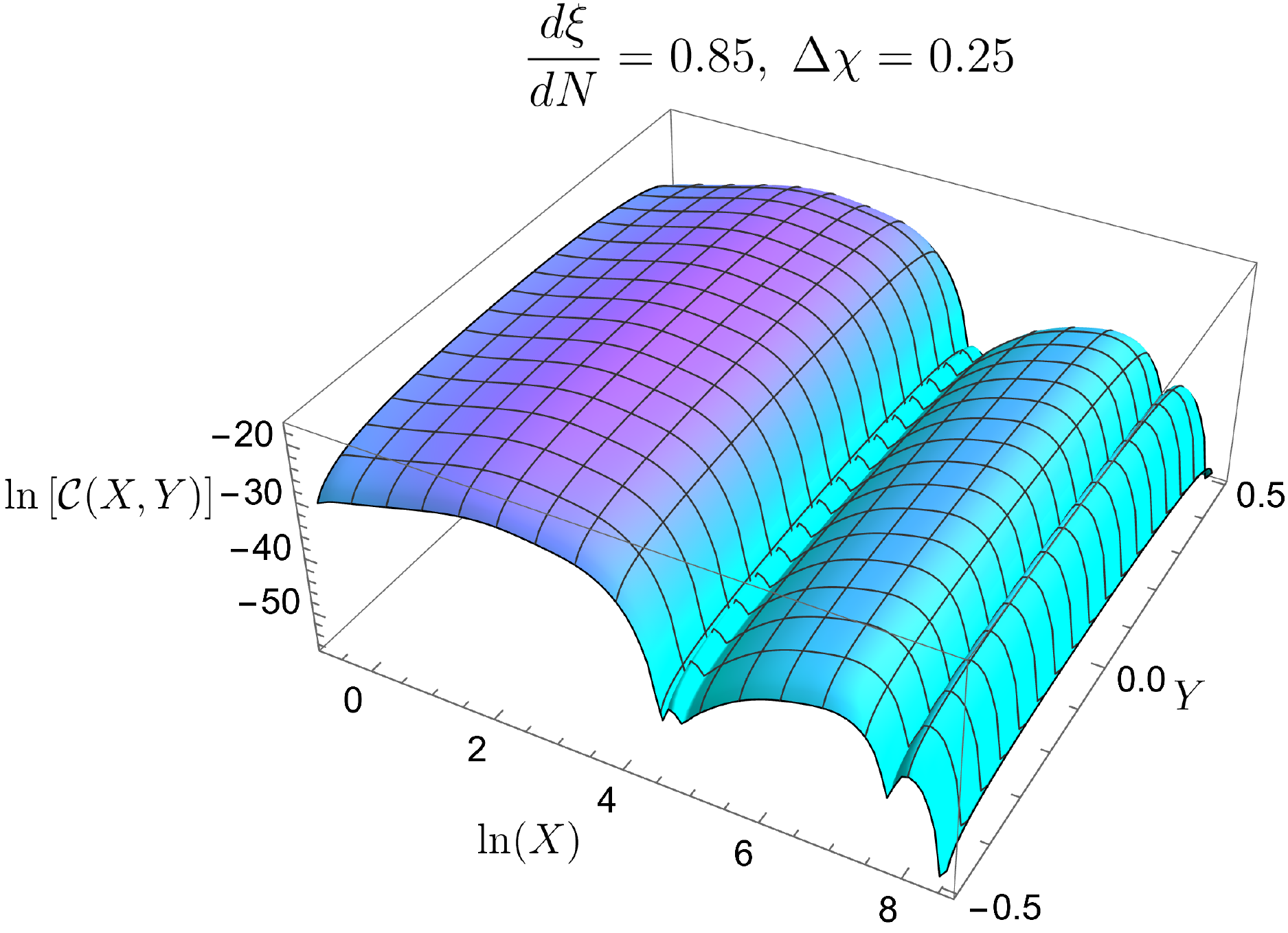}
    \includegraphics[width=0.45\linewidth]{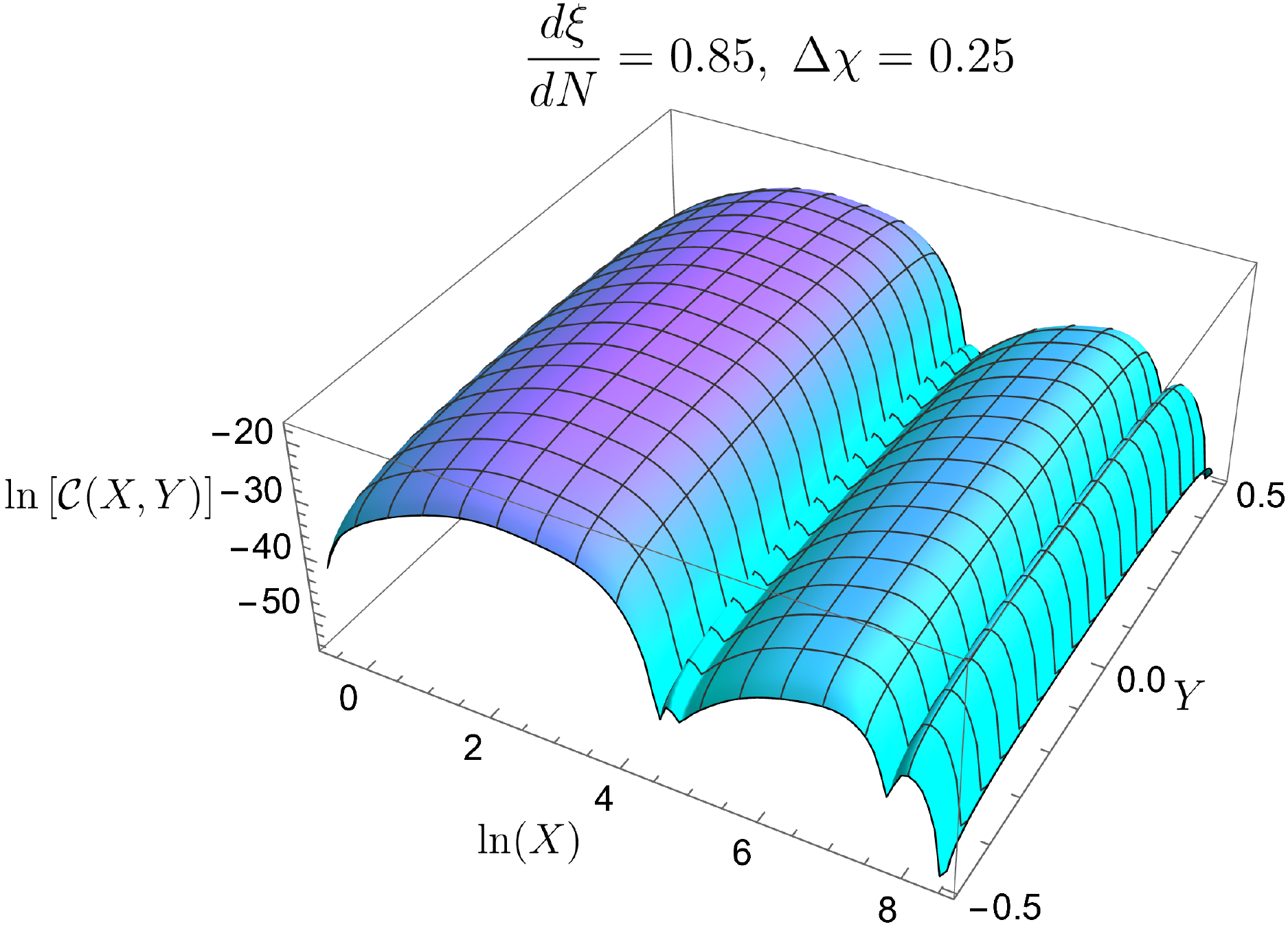}

    \includegraphics[width=0.45\linewidth]{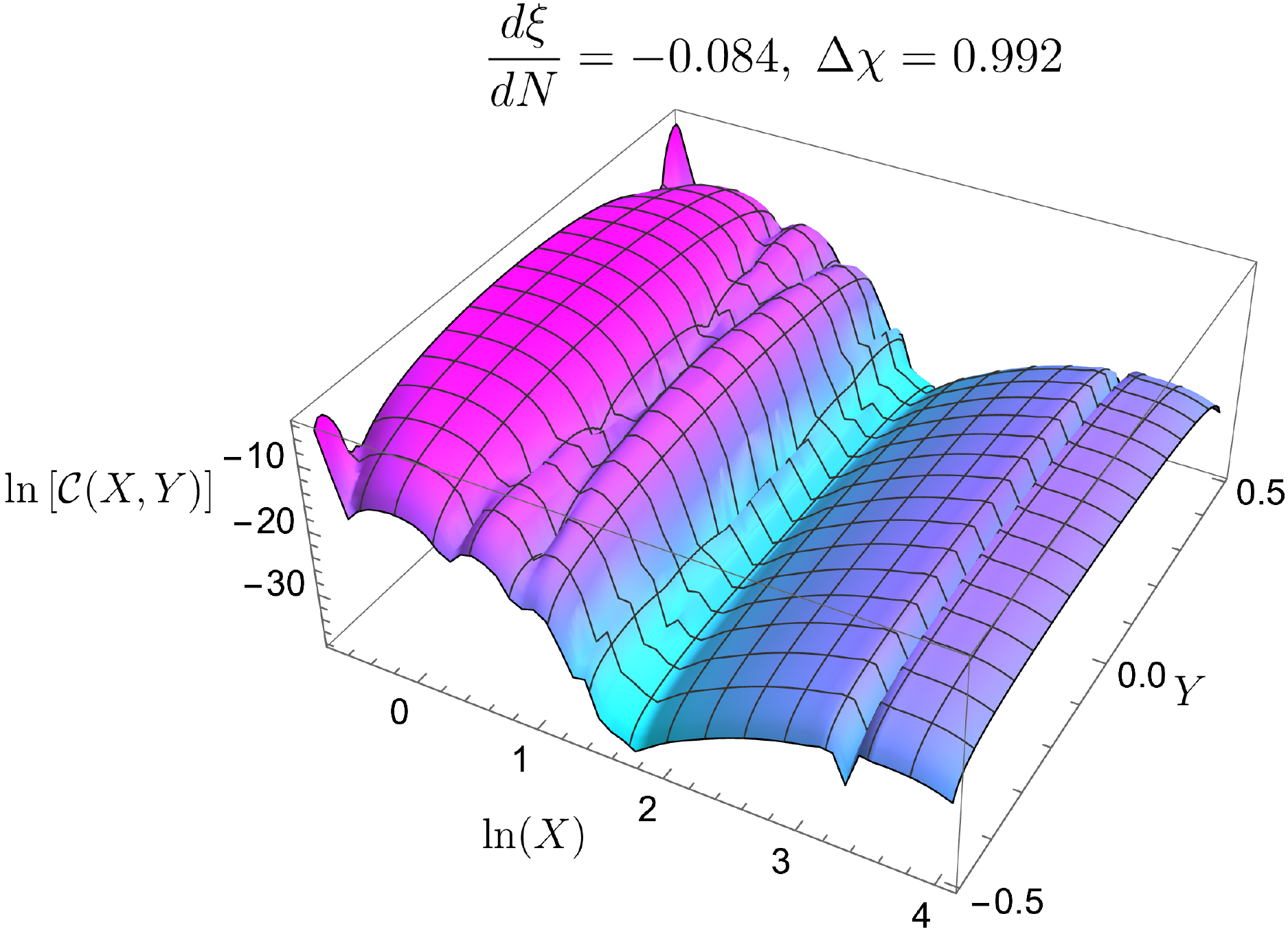}
    \includegraphics[width=0.45\linewidth]{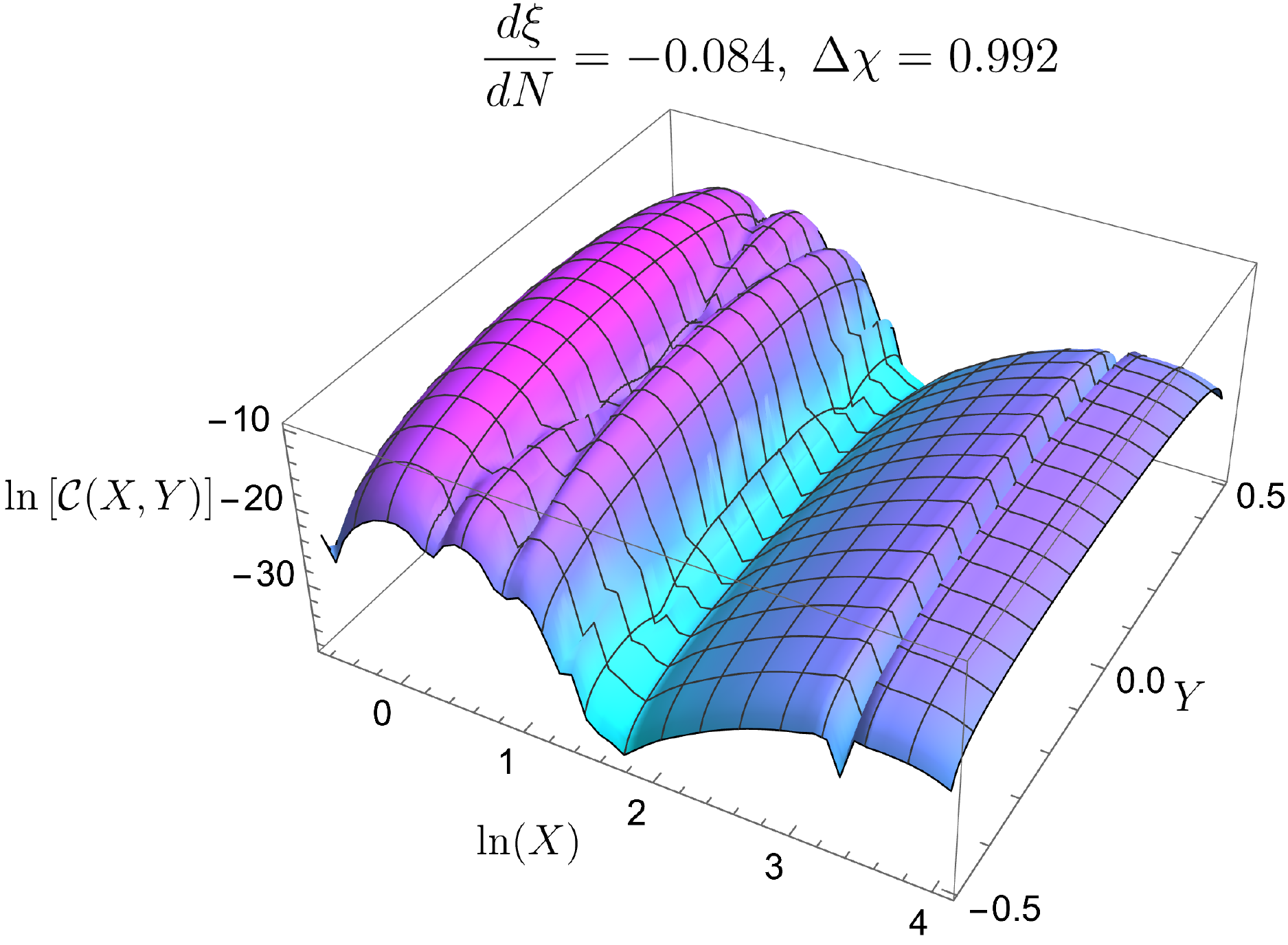}

    \includegraphics[width=0.45\linewidth]{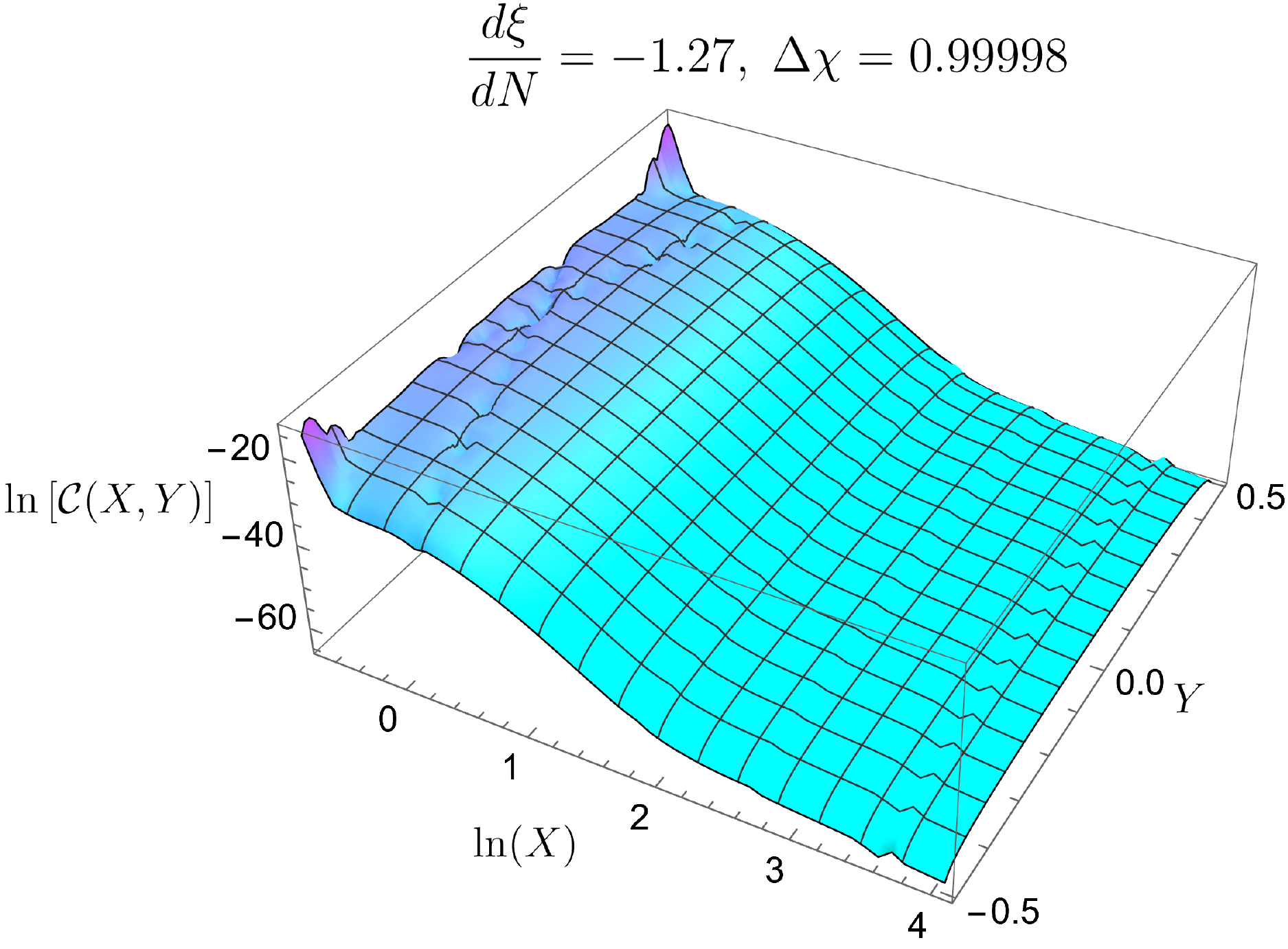}
    \includegraphics[width=0.45\linewidth]{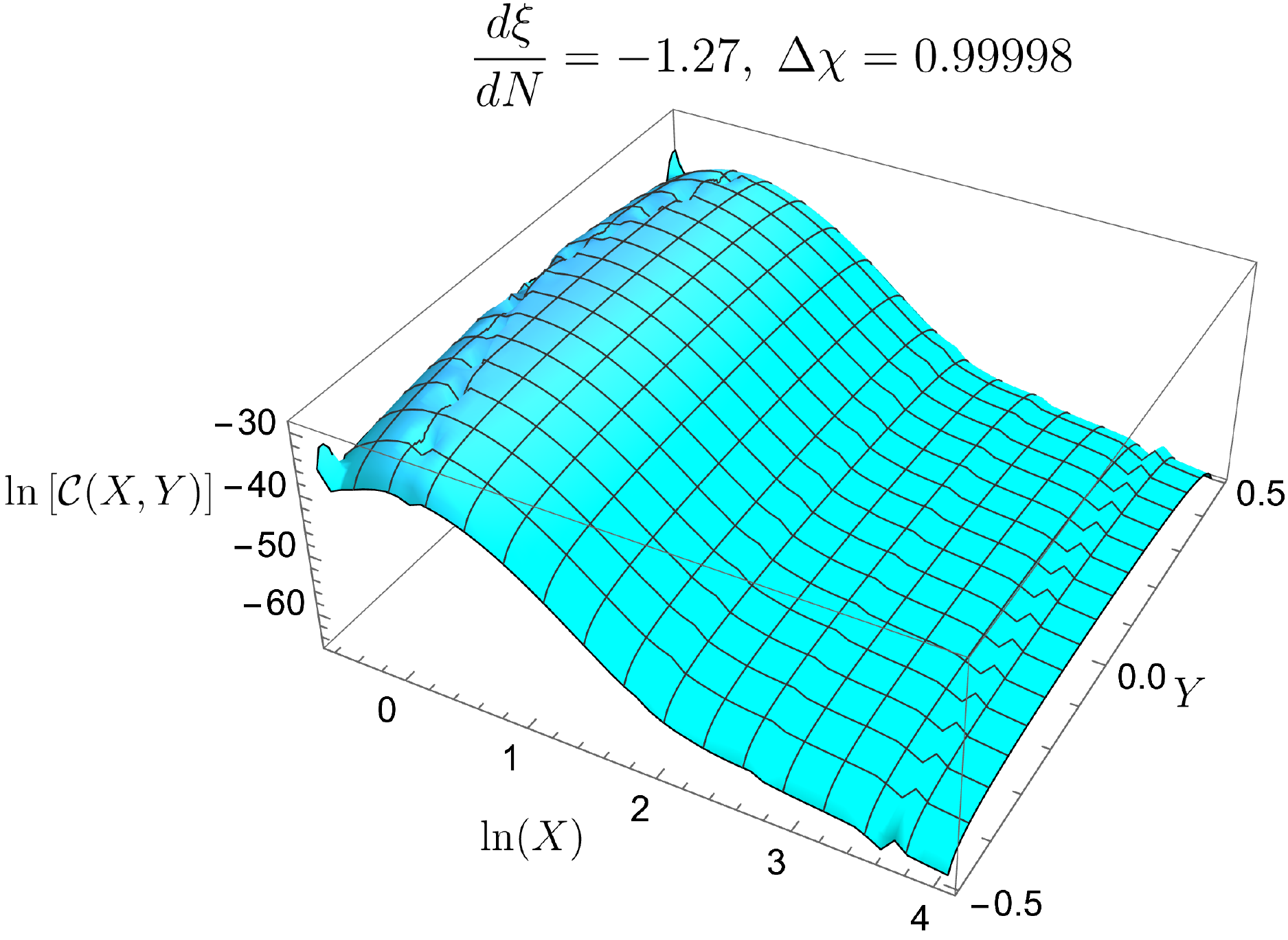}
    \caption{Three dimensional plots of the natural logarithm of the integrand defined in eq.~(\ref{code-N-PS}). The left side corresponds to the integrand of the plus polarization while the right side corresponds to the minus polarization. From top to bottom the panels correspond to the pair of squares, stars and triangles as they appear in figure \ref{fig:final-spectrum}. The instantaneous slope of the particle production parameter $\xi$ at the moment when the mode corresponding to the gravitational wave crossed the horizon as well as the chirality parameter defined in (\ref{eq:chirality}) is given for each panel.}
    \label{fig:integrand-parity}
\end{figure}

In this appendix we show the momentum dependence of the GW source, to support the considerations that we made in the final part of Section \ref{sec:results}. We recall that a GW of momentum $\vec{k}$ is sourced by two gauge modes, of momenta $\vec{p}$ and $\vec{q}$, that combine to give $\vec{p} + \vec{q} = \vec{k}$. In Figure \ref{fig:integrand-parity} we show the natural logarithm of the integrand ${\cal C}(X,Y)$, defined in (\ref{code-N-PS}), controlling the production, as a function of the two momenta. We show it in terms of the  variables defined in eq.~(\ref{pth2XY}). We recall that the region $X \gg 1/2$ (covering the outmost right part of each panel in the figure) corresponds to modes with $p,\, q \gg k$, which are anti-aligned so to add up to $\vec{k}$. The opposite regime $X \simeq 1/2$ corresponds to modes with $p ,\, q \la k$. The two points at the top left and bottom left of each panel correspond to the ``corners'' $X = 1/2$ and $Y = \pm 1/2$ of the integration domain, where one between $\vec{p}$ or $\vec{q}$ is equal to $\vec{k}$ while the other momentum vanishes. 

Figure~\ref{fig:integrand-parity} presents three rows, consisting of two panels each. The top, middle and bottom row corresponds, respectively, to the times marked with squares, stars, and triangles in Figure~\ref{fig:final-spectrum}. For each row, the left (resp., right) panel shows the integrand of eq.~(\ref{code-N-PS}) for the left-handed (resp., right-handed) GW polarization. From top to bottom the comoving momentum $\tilde{k}$ is $2.43\cdot 10^8$, $2.86\cdot 10^{11}$ and $3.09\cdot 10^{16}$ respectively. In each panel, we also indicate the value of $\frac{d \xi}{d N}$ at that moment, and the resulting amount of GW polarization, defined in eq.~(\ref{eq:chirality})

In the top row of the figure we observe a relatively small amount of polarization. At this time $\xi$ is rapidly increasing and so the GW production of the GW modes that leave the horizon around this time is dominated by gauge fields of higher momentum, that leave the horizon a few e-folds later, when $\xi$ is greater. This effect is eventually stopped by the phase space suppression, and we find that the region with $X\simeq {\cal O}(10)$ dominates the integral. For these values that gauge momenta are nearly anti-aligned, resulting in a polarization-independent production.~\footnote{Mathematically, we observe from eq.~(\ref{code-N-PS}) that, in the $X \gg 1/2$ limit the only terms carrying a dependence on the polarization actually becomes polarization-independent, namely $\left( 1 + 2 \lambda X \right)^4 \to 16 X^4$.} 

The second row (resp., third row) corresponds to times in which $\xi$ is nearly constant (resp., rapidly decreasing) both in an instantaneous sense and for some e-folds across the time shown. We observe from the figure that in the second row the production of the left-handed polarization is dominated by $X\simeq 1/2$, $Y\simeq 0$, namely by $p \simeq q \simeq \frac{k}{2}$. Among the three rows, this case is the one that resembles the most the result obtained at constant $\xi$. We see from the figure, and from the $\left( 1 + 2 \lambda X \right)^4$ factor in eq.~(\ref{code-N-PS}), that this region does not contribute to the right-handed GW polarization, hence explaining the significantly higher value of $\Delta \chi$ with respect to the top row. In the third row, characterized by a rapid decrease of $\xi$, we are in a situation in which the gauge modes whose wavelengths crossed the horizon long before the wavelength of the gravitational wave being produced are far more abundant than gauge modes of similar wavelength. This implies that the gauge mode dependence is heavily skewed towards the infrared, which is precisely the point at which the parity violation becomes maximal. This is manifest in the last row of panels in figure \ref{fig:integrand-parity} where the dominant contribution is consolidated in the upper and left-most corners of the integrand.

\end{document}